\documentclass[]{aa}
\usepackage{epsfig}
\def\lsim{\vcenter{\hbox{$<$}\offinterlineskip\hbox{$\sim$}}}
\def\gsim{\vcenter{\hbox{$>$}\offinterlineskip\hbox{$\sim$}}}
\begin{document}
\title{Dust-enshrouded giants in clusters in the Magellanic Clouds}
\author{Jacco Th. van Loon\inst{1}, Jonathan R. Marshall\inst{1},
        Albert A. Zijlstra\inst{2}}
\institute{Astrophysics Group, School of Physical \& Geographical Sciences,
           Keele University, Staffordshire ST5 5BG, UK
      \and School of Physics and Astronomy, University of Manchester,
           Sackville Street, P.O.Box 88, Manchester M60 1QD, UK}
\offprints{\email{jacco@astro.keele.ac.uk}}
\date{Received date; accepted date}
\titlerunning{Dust-enshrouded giants in clusters in the Magellanic Clouds}
\authorrunning{van Loon, Marshall, Zijlstra}
\abstract{We present the results of an investigation of post-Main Sequence
mass loss from stars in clusters in the Magellanic Clouds, based around an
imaging survey in the L$^\prime$-band (3.8 $\mu$m) performed with the VLT at
ESO. The data are complemented with JHK$_{\rm s}$ (ESO and 2MASS) and mid-IR
photometry (TIMMI2 at ESO, ISOCAM on-board ISO, and data from IRAS and MSX).
The goal is to determine the influence of initial metallicity and initial mass
on the mass loss and evolution during the latest stages of stellar evolution.
Dust-enshrouded giants are identified by their reddened near-IR colours and
thermal-IR dust excess emission. Most of these objects are Asymptotic Giant
Branch (AGB) carbon stars in intermediate-age clusters, with progenitor masses
between 1.3 and $\sim$5 M$_\odot$. Red supergiants with circumstellar dust
envelopes are found in young clusters, and have progenitor masses between 13
and 20 M$_\odot$. Post-AGB objects (e.g., Planetary Nebulae) and massive stars
with detached envelopes and/or hot central stars are found in several
clusters. We model the spectral energy distributions of the cluster IR
objects, in order to estimate their bolometric luminosities and mass-loss
rates. The IR objects are the most luminous cluster objects, and have
luminosities as expected for their initial mass and metallicity. They
experience mass-loss rates in the range from a few $10^{-6}$ up to $10^{-4}$
M$_\odot$ yr$^{-1}$ (or more), with most of the spread being due to
evolutionary effects and only a weak dependence on progenitor mass and/or
initial metallicity. About half of the mass lost by 1.3--3 M$_\odot$ stars is
shed during the superwind phase, which lasts of order $10^5$ yr. Objects with
detached shells are found to have experienced the highest mass-loss rates, and
are therefore interpreted as post-superwind objects. We also propose a simple
method to measure the cluster mass from L$^\prime$-band images.
}
\maketitle

\section{Introduction}

Near the end of their life, stars of initial mass $M_{\rm i}\sim$1--8
M$_\odot$ and many of the more massive stars become cool and luminous giants,
through evolution along the first ascent Red Giant Branch (RGB) and Asymptotic
Giant Branch (AGB), or as a red supergiant (RSG), respectively. As such, they
become a powerful tracer of the underlying stellar population in clusters and
galaxies. In these final stages of evolution the red giant loses a significant
fraction of its mass at rates of $\dot{M}\sim10^{-6}$ to $10^{-3}$ M$_\odot$
yr$^{-1}$ in a "superwind" driven as a result of stellar pulsation and
circumstellar dust formation. These stars thus enrich the interstellar medium
with dust and products of nuclear burning on timescales from $\sim$10 Myr for
the most massive RSGs up to $\sim$10 Gyr for stars like the Sun.

Progress in the theoretical understanding of evolution and mass loss of red
giants is hampered by the difficulty to empirically chart the results of
stellar evolution (e.g., luminosity, mass-loss rate, chemical abundances) onto
the boundary conditions of initial mass and initial metallicity. It is very
difficult to measure the mass of a red giant and its metallicity, let alone
its {\it initial} mass and metallicity. A powerful way around this problem is
to study red giants in clusters for which the age (and hence the mass of the
red giant) and initial metallicity can often be obtained with reasonable ease
and accuracy.

Studies of clusters in the Magellanic Clouds (Mould \& Aaronson 1979;
Lloyd-Evans 1980, 1983; Frogel \& Cohen 1982; Aaronson \& Mould 1985; Frogel,
Mould \& Blanco 1990; Ferraro et al.\ 1995; Marigo, Girardi \& Chiosi 1996)
confirm some of the main features of stellar evolution models for AGB stars,
such as the transition from oxygen-rich M stars, through S stars when carbon
and oxygen are equally abundant, into carbon stars. In each cluster, the
carbon stars are more luminous than the M stars, with the S stars in between.
The transition luminosity M-S and S-C is higher in younger and/or more
metal-rich clusters. These studies could not reach the phases of high mass
loss, where circumstellar reddening becomes important. Observations at
thermal-IR wavelengths to detect circumstellar dust emission are essential to
measure the mass-loss rate from these stars.

Only a few cluster IR objects are known. These include the $M_{\rm i}\sim4$
M$_\odot$ OH/IR star IRAS\,05298$-$6957 (van Loon et al.\ 2001a) and the
$M_{\rm i}\sim2.2$ M$_\odot$ carbon star LI-LMC\,1813 (van Loon et al.\ 2003).
Tanab\'{e} et al.\ (2004) surveyed clusters in the Magellanic Clouds with the
ISOCAM instrument onboard ISO. They detected two dust-enshrouded AGB stars in
each of the populous intermediate-age clusters NGC\,419 (SMC) and NGC\,1783
and NGC\,1978 (LMC), four of which were already known from near-IR data.

We here present results from a systematic survey for dust-enshrouded red
giants in clusters in the Magellanic Clouds. The investigation is built around
L$^\prime$-band imaging obtained at the ESO-VLT, supplemented by near-IR
imaging photometry and a collection of mid-IR data from IRAS, ISO, MSX and the
ESO 3.6m telescope. We find 30 dusty stars, of which 26 are likely cluster
members. We determine their nature, luminosity and mass-loss rate, and
investigate the dependence on progenitor mass and metallicity.

\section{Cluster selection, properties, and bias}

We have attempted to select clusters across a wide range of ages and
metallicities (Table 1, criterion ``C''). Although the more populous clusters
are attractive targets to survey for rare objects such as dust-enshrouded
stars, small clusters of $t\lsim1$ Gyr are much more numerous and many of the
IR-object selected clusters (Table 1, criterion ``I'') turn out to be
associated with this type of cluster. The drawback of these less conspicuous
clusters is that they have not usually been studied in detail and ages and
metallicities are often inaccurate --- if at all available. Due to the smaller
stellar over-density with respect to the field, membership of small clusters
is also more ambiguous. We added from the literature (Table 1, criterion
``L'') another two clusters in which IR objects had been found.

The cluster ages and metallicities --- for which we take the commonly used
relative iron abundance [Fe/H] --- are collected from a variety of recent
sources in the literature, and are listed in Table 1. For some clusters,
average values were taken from multiple sources. In the case of NGC\,1978 the
disagreement between the individual sources is rather large, with Olszewski et
al.\ (1991) and Hill et al.\ (2000) disagreeing over the metallicity by 0.5
dex for the spectroscopic determinations of two stars in common.

%
%
\begin{figure}[]
\centerline{\psfig{figure=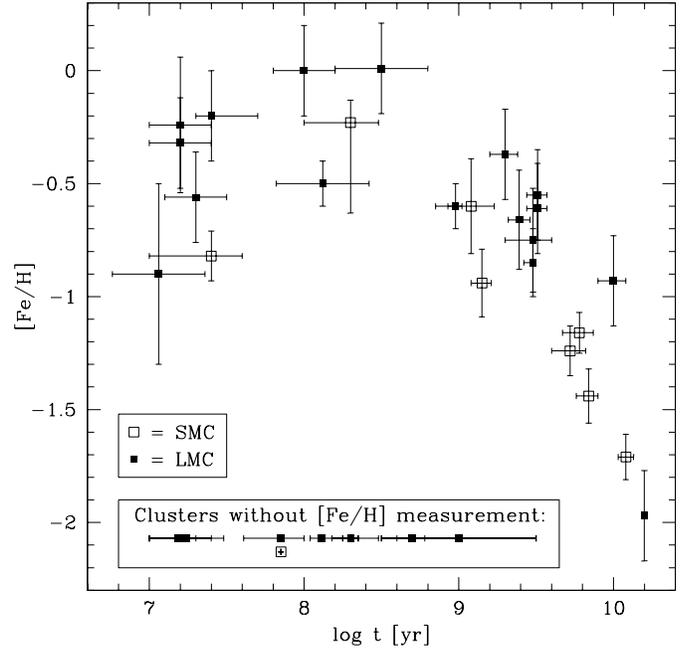,width=88mm}}
\caption[]{Age-metallicity diagram for the SMC (open symbols) and LMC (solid
symbols) clusters in our sample.}
\end{figure}

The clusters show a clear correlation between age and metallicity (Fig.\ 1),
which is observed for both the SMC and the LMC (e.g., Olszewski et al.\ 1991;
Geisler et al.\ 1997; Da Costa \& Hatzidimitriou 1998; de Freitas Pacheco,
Barbuy \& Idiart 1998). The turnover to lower metallicities of the youngest
clusters is intriguing, with young, fairly massive clusters as metal-poor as
[Fe/H]$=-0.9$ observed in both clouds (Hill et al.\ 2000). There is sufficient
spread in the cluster properties that one cannot assume that every LMC cluster
is more metal-rich than an SMC cluster of the same age.

%
%
\begin{figure}[]
\centerline{\psfig{figure=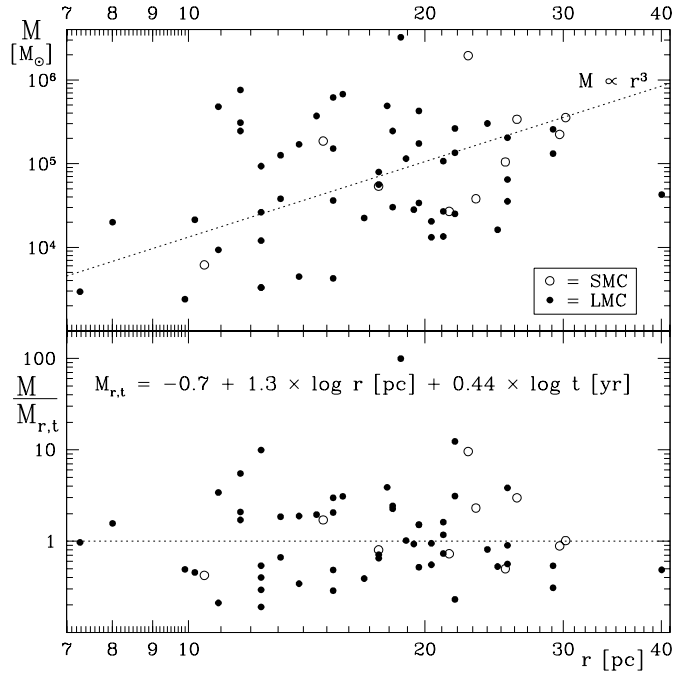,width=88mm}}
\caption[]{Mass-radius diagram for clusters in the SMC (open symbols) and LMC
(solid symbols), with masses from Mackey \& Gilmore (2003a,b) and radii from
Bica \& Schmitt (1995) and Bica et al.\ (1999). The bottom panel shows the
mass ratio with respect to the displayed parameterization $M_{r,t}$.}
\end{figure}

%
%
\begin{table*}
\caption[]{List of programme clusters, in order of increasing Right Ascension
(all coordinates are in J2000). The cluster radii are taken from Bica \&
Schmitt (1995) and Bica et al.\ (1999) for the SMC and LMC, respectively, and
converted to parsecs assuming distances to the SMC and LMC of 60 and 50 kpc,
respectively. Metallicities without error estimates are assumed values; masses
without error estimates are derived from the mass-radius-age relation as
described in the text. Also given are the numbers of stars detected in the
L$^\prime$-band images with an L$^\prime$-band brightness exceeding that of
the RGB tip, at $M_{{\rm L}^\prime}<-6.4$ mag (see Section 5.5). The
references for the ages, metallicities and masses are as follows:
 1=Beasley, Hoyle \& Sharples (2002);
 2=Bica et al.\ (1996);
 3=de Freitas Pacheco et al.\ (1998);
 4=Elson (1991);
 5=Kontizas (1980);
 6=Mackey \& Gilmore (2003a);
 7=Mackey \& Gilmore (2003b);
 8=Mateo (1988);
 9=Mighell, Sarajedini \& French (1998);
10=Oliva \& Origlia (1998);
11=Olszewski et al.\ (1991);
12=Pietrzy\'{n}ski \& Udalski (2000);
13=Vallenari, Bettoni \& Chiosi (1998);
14=van Loon et al.\ (2001a);
15=van Loon et al.\ (2003);
 x=average of de Freitas Pacheco et al.\ (1998), Hill et al.\ (2000) \&
   Olszewski et al.\ (1991);
 y=this work.
The last column indicates the criterion upon which the selection was based
(C=''cluster'', I=''IR'', L=``Literature'').}
\begin{tabular}{lcccclllccc}
\hline\hline
Cluster                                &
RA                                     &
Dec                                    &
r                                      &
r                                      &
\multicolumn{1}{c}{log(t)}             &
\multicolumn{1}{c}{[Fe/H]}             &
\multicolumn{1}{c}{log(M)}             &
N$_{> \rm RGB}$                        &
Ref                                    &
Sel                                    \\
                                       &
($^{\rm h}$ $^{\rm m}$ $^{\rm s}$)     &
($^\circ$ $^\prime$ $^{\prime\prime}$) &
( $^\prime$ )                          &
(pc)                                   &
\multicolumn{1}{c}{(yr)}               &
                                       &
\multicolumn{1}{c}{(M$_\odot$)}        &
                                       &
                                       &
                                       \\
\hline
\multicolumn{11}{l}{\it Small Magellanic Cloud} \\
ESO\,28-19            &
0 24 46.0             &
$-$72 47 37           &
1.70                  &
\llap{2}9.7           &
$9.78^{+0.09-0.11}$   &
$-1.16\pm0.09$        &
$5.35\pm0.12$         &
1                     &
7                     &
C                     \\
NGC\,121              &
0 26 49.0             &
$-$71 32 10           &
1.73                  &
\llap{3}0.2           &
\llap{1}$0.08\pm0.05$ &
$-1.71\pm0.10$        &
$5.55\pm0.10$         &
1                     &
7                     &
C                     \\
NGC\,152              &
0 32 56.0             &
$-$73 06 59           &
1.50                  &
\llap{2}6.2           &
$9.15^{+0.06-0.07}$   &
$-0.94\pm0.15$        &
$5.53^{+0.52-0.85}$   &
5                     &
7                     &
C                     \\
NGC\,330              &
0 56 20.0             &
$-$72 27 44           &
1.33                  &
\llap{2}3.2           &
$7.40^{+0.20-0.40}$   &
$-0.82\pm0.11$        &
$4.58^{+0.20-0.18}$   &
15                    &
7                     &
C                     \\
HW\,32                &
0 57 24.0             &
$-$71 11 00           &
0.33                  &
5.8                   &
\llap{$\lsim$}7.85    &
$-0.7$                &
3.7                   &
0                     &
5                     &
C                     \\
NGC\,416              &
1 07 58.0             &
$-$72 21 25           &
0.85                  &
\llap{1}4.8           &
$9.84^{+0.06-0.08}$   &
$-1.44\pm0.12$        &
$5.27^{+0.18-0.17}$   &
2                     &
7                     &
C                     \\
NGC\,419              &
1 08 19.5             &
$-$72 53 03           &
1.40                  &
\llap{2}4.4           &
$9.08^{+0.15-0.23}$   &
$-0.60\pm0.21$        &
5.1                   &
23                    &
3                     &
C                     \\
NGC\,458              &
1 14 54.0             &
$-$71 32 58           &
1.23                  &
\llap{2}1.5           &
$8.30^{+0.18-0.30}$   &
$-0.23^{+0.1-0.4}$    &
$4.43^{+0.29-0.27}$   &
0                     &
7                     &
C                     \\
ESO\,30-4             &
1 49 30.0             &
$-$73 43 40           &
2.20                  &
\llap{3}8.4           &
$9.72^{+0.10-0.12}$   &
$-1.24\pm0.11$        &
5.6                   &
1                     &
9                     &
C                     \\
\multicolumn{11}{l}{\it Large Magellanic Cloud} \\
NGC\,1651             &
4 37 31.1             &
$-$70 35 02           &
1.35                  &
\llap{1}9.6           &
$9.30^{+0.08-0.10}$   &
$-0.37\pm0.20$        &
$5.24^{+0.45-0.28}$   &
1                     &
6                     &
C                     \\
HS\,33                &
4 49 17.0             &
$-$69 52 40           &
0.40                  &
5.8                   &
$8.11^{+0.19-0.07}$   &
$-0.4$                &
3.9                   &
4                     &
y                     &
I                     \\
KMHK\,29\rlap{2}      &
4 55 35.0             &
$-$69 26 50           &
0.53                  &
7.7                   &
$7.18^{+0.12-0.18}$   &
$-0.4$                &
3.6                   &
3                     &
y                     &
I                     \\
KMHK\,28\rlap{5}      &
4 55 39.0             &
$-$67 49 11           &
0.29                  &
4.2                   &
\llap{$\gsim$}9.00    &
$-0.4$                &
4.1                   &
1                     &
y                     &
I                     \\
NGC\,1783             &
4 59 08.0             &
$-$65 59 18           &
2.50                  &
\llap{3}6.4           &
$9.48^{+0.12-0.18}$   &
$-0.75\pm0.23$        &
5.5                   &
-                     &
3                     &
L                     \\
NGC\,1818             &
5 04 13.8             &
$-$66 26 02           &
1.45                  &
\llap{2}1.1           &
$7.40^{+0.30-0.10}$   &
$-0.20\pm0.20$        &
$4.13^{+0.15-0.14}$   &
13                    &
6                     &
C                     \\
NGC\,1831             &
5 06 17.4             &
$-$64 55 11           &
1.75                  &
\llap{2}5.5           &
$8.50\pm0.30$         &
$+0.01\pm0.20$        &
$4.81\pm0.13$         &
4                     &
6                     &
C                     \\
NGC\,1852             &
5 09 23.0             &
$-$67 46 42           &
0.95                  &
\llap{1}3.8           &
$9.48^{+0.01-0.06}$   &
$-0.85\pm0.15$        &
5.0                   &
6                     &
1                     &
I                     \\
NGC\,1866             &
5 13 38.9             &
$-$65 27 52           &
2.75                  &
\llap{4}0.0           &
$8.12\pm0.30$         &
$-0.50\pm0.10$        &
$4.63\pm0.08$         &
8                     &
6                     &
C                     \\
SL\,349               &
5 16 54.4             &
$-$68 52 36           &
0.46                  &
6.7                   &
$8.70^{+0.08-0.10}$   &
$-0.4$                &
4.2                   &
0                     &
13                    &
I                     \\
NGC\,1903             &
5 17 22.4             &
$-$69 20 16           &
0.95                  &
\llap{1}3.8           &
$7.85^{+0.15-0.24}$   &
$-0.4$                &
4.2                   &
19                    &
13                    &
I                     \\
HS\,270               &
5 23 52.0             &
$-$71 34 42           &
0.60                  &
8.7                   &
$9.00\pm0.5$          &
$-0.4$                &
4.5                   &
1                     &
y                     &
I                     \\
SL\,453               &
5 25 01.1             &
$-$69 26 03           &
0.63                  &
9.2                   &
$8.30\pm0.05$         &
$-0.4$                &
4.2                   &
4                     &
11                    &
I                     \\
SL\,482               &
5 27 17.9             &
$-$66 22 06           &
0.34                  &
4.9                   &
$7.24\pm0.24$         &
$-0.4$                &
3.4                   &
1                     &
2                     &
I                     \\
NGC\,1984             &
5 27 40.8             &
$-$69 08 05           &
0.68                  &
9.9                   &
$7.06\pm0.30$         &
$-0.90\pm0.40$        &
$3.38^{+0.35-0.28}$   &
6                     &
6                     &
C                     \\
BSDL\,183\rlap{7}     &
5 27 51.0             &
$-$69 13 20           &
0.30                  &
4.4                   &
$9.00\pm0.5$          &
$-0.4$                &
4.1                   &
4                     &
y                     &
I                     \\
NGC\,1994             &
5 28 22.0             &
$-$69 08 30           &
0.78                  &
\llap{1}1.3           &
$7.20\pm0.20$         &
$-0.24\pm0.30$        &
3.8                   &
5                     &
10                    &
I                     \\
NGC\,1978             &
5 28 45.0             &
$-$66 14 14           &
1.68                  &
\llap{2}4.4           &
$9.39\pm0.07$         &
$-0.66\pm0.22$        &
5.2                   &
18                    &
x                     &
C                     \\
HS\,327-E             &
5 29 23.3             &
$-$69 55 10           &
0.45                  &
6.5                   &
$8.30^{+0.18-0.12}$   &
$-0.4$                &
4.0                   &
-                     &
14                    &
L                     \\
SL\,519               &
5 30 01.7             &
$-$69 57 02           &
0.33                  &
4.8                   &
$8.30\pm0.05$         &
$-0.4$                &
3.8                   &
8                     &
12                    &
I                     \\
NGC\,2002             &
5 30 21.0             &
$-$66 53 02           &
0.90                  &
\llap{1}3.1           &
$7.20\pm0.20$         &
$-0.3$                &
3.9                   &
8                     &
4                     &
I                     \\
NGC\,2010             &
5 30 34.6             &
$-$70 49 10           &
0.90                  &
\llap{1}3.1           &
$8.00\pm0.20$         &
$-0.00\pm0.20$        &
4.3                   &
0                     &
8                     &
C                     \\
NGC\,2004             &
5 30 40.9             &
$-$67 17 09           &
1.45                  &
\llap{2}1.1           &
$7.30\pm0.20$         &
$-0.56\pm0.20$        &
$4.43^{+0.24-0.23}$   &
6                     &
6                     &
I                     \\
NGC\,2100             &
5 42 08.6             &
$-$69 12 44           &
1.25                  &
\llap{1}8.2           &
$7.20\pm0.20$         &
$-0.32\pm0.20$        &
$4.48^{+0.33-0.30}$   &
11                    &
6                     &
I                     \\
NGC\,2121             &
5 48 11.6             &
$-$71 28 51           &
1.23                  &
\llap{1}7.9           &
$9.51^{+0.06-0.07}$   &
$-0.61\pm0.20$        &
$5.69^{+0.22-0.44}$   &
4                     &
6                     &
C                     \\
NGC\,2155             &
5 58 33.3             &
$-$65 28 35           &
1.20                  &
\llap{1}7.5           &
$9.51^{+0.06-0.07}$   &
$-0.55\pm0.20$        &
$4.90^{+0.26-0.23}$   &
2                     &
6                     &
C                     \\
ESO\,121-3            &
6 02 02.0             &
$-$60 31 20           &
1.05                  &
\llap{1}5.3           &
\llap{1}$0.00^{+0.08-0.10}$ &
$-0.93\pm0.20$        &
5.2                   &
0                     &
11                    &
C                     \\
KMHK\,1603            &
6 02 30.0             &
$-$67 12 54           &
0.75                  &
\llap{1}0.9           &
$8.98^{+0.04-0.05}$   &
$-0.60\pm0.10$        &
4.6                   &
2                     &
15                    &
I                     \\
NGC\,2210             &
6 11 31.5             &
$-$69 07 17           &
1.65                  &
\llap{2}4.0           &
\llap{1}$0.20\pm0.01$ &
$-1.97\pm0.20$        &
$5.48\pm0.10$         &
0                     &
6                     &
C                     \\
\hline
\end{tabular}
\end{table*}

Mackey \& Gilmore (2003a,b) estimated the cluster masses for 10 clusters in
the SMC and 53 clusters in the LMC, some of which are in common with our
sample. We collected the radii for all their clusters from Bica \& Schmitt
(1995) and Bica et al.\ (1999). We thus find a weak correlation between the
mass and volume (Fig.\ 2, top panel):
\begin{equation}
\log M_r [{\rm M}_\odot] = 1.12\ +\ 3 \log r [{\rm pc}],
\end{equation}
with a standard deviation of 0.7 in $\log M$. The scatter can be reduced by a
parameterization in both radius and cluster age (Fig.\ 2, bottom panel):
\begin{equation}
\log M_{r,t} [{\rm M}_\odot] = a\ +\ b \log r [{\rm pc}]\ +\ c \log t [{\rm
yr}],
\end{equation}
where $a=-0.7\pm0.5$, $b=1.3\pm0.4$ and $c=0.44\pm0.06$. We use the $M_{r,t}$
relation to estimate the mass of clusters that were not included in the
analysis of Mackey \& Gilmore (Table 1).

Throughout this paper we use distance moduli to the SMC and LMC of 19.0 and
18.5 mag, respectively.

\section{Observations}

\subsection{J and K$_{\rm s}$-band imaging}

Images were obtained through the J (1.25 $\mu$m) and K$_{\rm s}$ (2.2 $\mu$m)
bands with the SOFI instrument on the ESO 3.5m NTT at La Silla, Chile, on the
first half of the night of 7 November 2000. At a scale of
$0.146^{\prime\prime}$ pixel$^{-1}$, the field-of-view was
$2.5^\prime\times2.5^\prime$. The images were constructed through combination
of 9 frames which were obtained at slightly different (few arcsec) "jitter"
positions to allow for the construction and subtraction of the (bright and
variable) sky background. The seeing was typically around 1$^{\prime\prime}$,
but some of the images were distorted because the delay time to allow the
telescope to recover from its movement to a new jitter position was set too
short. The weather conditions deteriorated shortly after completion of the
cluster-selected subsample, and consequently the J and K$_{\rm s}$-band data
obtained for the IR-object-selected subsample are often quite poor and in
several cases not obtained at all. The total integration time per final image
is 90 seconds except where images were discarded because no stars were visible
due to cloud cover.

The data were reduced using the ESO Eclipse and MIDAS software packages.
First, electronic ghosts were removed with help of the Eclipse tool {\sc
is\_ghost}. Then, a flatfield frame was constructed from the differences
between the pair-wise observations of an illuminated and dark screen, to
correct for transmission, vignetting and detector response variations across
the frame. Next, each frame was sky-subtracted by first (for each pixel)
subtracting the minimum value of the set of 9 jittered frames (after having
scaled each frame to its median value), and then subtracting the pixel value
obtained by median-averaging along the columns of the frame. The resulting
frames were shift-added to create the final image.

Photometry was obtained for all the stars detected in the L$^\prime$-band
images described below, using (software) apertures with a diameter of
$1.6^{\prime\prime}$. A few stars in NGC\,419, NGC\,1984 and NGC\,2210 were
measured through $0.7^{\prime\prime}$-diameter apertures in the case of
exceptional crowding, and their magnitudes were corrected for the aperture
difference. The magnitudes were calibrated against those of known 2MASS
sources in the images to a systematic accuracy of $\sim$0.1 mag, but relative
photometry within the images themselves is generally better.

2MASS data are used where ESO J and K$_{\rm s}$-band data are absent, as well
as to complement the analysis with H-band (1.65 $\mu$m) photometry. The J-band
magnitude of NGC\,1994-IR1 and the K-band magnitudes of two other bright stars
in the same cluster were reconstructed by differential aperture photometry on
the 2MASS image. The 2MASS data are problematic for clusters in the Magellanic
Clouds though, as many 2MASS sources are in fact blends of stars. Also, the
2MASS data were obtained more than two years before the L$^\prime$-band data
whilst the ESO J and K$_{\rm s}$-band data were obtained within a few weeks
after the L$^\prime$-band data. Many of the dust-enshrouded red giants are
large-amplitude variables with periods of around 400--1000 days and hence,
where the ESO photometry can be considered near-simultaneous with the
L$^\prime$-band photometry, the 2MASS photometry have random phase differences
with respect to the L$^\prime$-band photometry.

Some objects have been detected in the DENIS survey of the Magellanic Clouds
(Cioni et al.\ 2000). Although less deep than 2MASS, it provides an I-band
magnitude for some of the (optically) brighter, relatively isolated objects.
This is useful in constraining the spectral energy distributions of hot stars
such as post-AGB objects and PNe. The DENIS catalogue also lists B and/or
R-band magnitudes in some cases. Crowding is more severe in the optical,
though, which could render the optical colours of intrinsically red objects
bluer due to the contribution at short wavelengths of other, less red stars.

\subsection{L$^\prime$-band imaging}

Images were obtained through the L$^\prime$ band (3.78 $\mu$m, bandwidth
${\Delta}\lambda=0.58$ $\mu$m) with the ISAAC instrument on the ESO VLT at
Paranal, Chile, in service mode during a total of three hours spread over the
nights of 12, 17 and 18 October 2000. At a scale of $0.071^{\prime\prime}$
pixel$^{-1}$, the field-of-view was
$72^{\prime\prime}\times72^{\prime\prime}$. The images were obtained following
the usual chop-and-nod procedure for thermal-IR observations, where both the
chop and the nod throw were $10^{\prime\prime}$ in the North-South direction.
The total exposure time per target is two minutes, with a detector integration
time of 0.104 sec. The conditions were photometric, with a seeing of typically
around $0.5^{\prime\prime}$.

The data were reduced using the ESO Eclipse and MIDAS software packages.
First, electronic ghosts were removed with help of the Eclipse tool {\sc
is\_ghost}. A flatfield frame was constructed by imaging the twilight sky and
using the Eclipse tool {\sc is\_twflat}. After flatfielding the science frames
obtained at each nod position, the resulting pair of frames were subtracted
and then shift-added to produce a frame with stellar images that contain all
of the observed light. Photometric measurements were obtained through a
$1.07^{\prime\prime}$ software aperture, and were calibrated against the
measurements of the standard stars HD\,22686, HD\,19904 and HR\,2354.

Additional photometry was obtained from acquisition images for L-band
spectroscopy for the clusters NGC\,1903, SL\,453, NGC\,1984, NGC\,1978 and
SL\,519 (7 \& 8 December 2003; van Loon et al., in preparation) and for
KMHK\,1603 (14 December 2001; van Loon et al.\ 2003). These images were
obtained through the L$^\prime$-band filter except for NGC\,1978 and the
standard star HIP\,020020\,A+B that were observed through narrow-band filters
centred at 3.8 and 3.28 $\mu$m, respectively. Detector integration times were
0.025 sec for the clusters and 0.28 sec for NGC\,1978 and the standard star.
The same zeropoints were assumed for all images, which by cross-correlation
with our dedicated L$^\prime$-band photometry seemed to be accurate to within
a few 0.1 mag.

\subsection{N-band imaging}

Images were obtained in the N band (8--13 $\mu$m) with the TIMMI2 instrument
on the ESO 3.6m at La Silla, Chile, on the nights of 19--21 January 2001, of
the clusters NGC\,330 and KMHK\,1603 through the N2 filter (central wavelength
10.6 $\mu$m, 9.7--11.3 $\mu$m at 50\% transmission) and of the clusters
NGC\,419, NGC\,1984 and NGC\,1994 through the 11.9 $\mu$m filter (11.0--12.2
$\mu$m at 50\% transmission). At a scale of $0.3^{\prime\prime}$ pixel$^{-1}$
($0.2^{\prime\prime}$ pixel$^{-1}$ for NGC\,330), the field-of-view was
$96^{\prime\prime}\times72^{\prime\prime}$
($64^{\prime\prime}\times48^{\prime\prime}$ for NGC\,330). Standard chopping
and nodding procedures were followed, with a throw of $10^{\prime\prime}$.
Total effective exposure times amounted to 4--9 minutes per cluster, using
detector integration times of 22 to 27 msec.

The data were reduced using the ESO MIDAS software package. After inspection
of all individual frames and rejection of saturated data, the frames belonging
to the same nod position were averaged and then subtracted from the average
frame at the complementary nod position. The thus obtained images were
shift-added to construct the final image, from which by additional
shift-adding an image was produced with stellar images that contain all the
observed light. On these stellar images photometry was obtained through a
$3^{\prime\prime}$ (software) aperture and which was calibrated against
measurements of the standard stars $\lambda_2$\,Tuc, $\alpha$\,Hya,
$\gamma$\,Pic, $\gamma$\,Ret and $\theta$\,Dor. An early result for
LI-LMC\,1813 (KMHK\,1603-IR1) has already been presented in van Loon et al.\
(2003).

\subsection{Additional mid-IR imaging}

For the identified cluster IR objects, additional mid-IR data was obtained
from the IRAS and MSX surveys and in some cases from ISO survey data. Like
2MASS, these data are not contemporary with the L$^\prime$-band data, however
they are important in constraining the spectral energy distribution to
estimate the bolometric luminosity and mass-loss rate.

For each IR object we collected the corresponding IRAS scans from the IRAS
data server\footnote{http://www.astro.rug.nl/IRAS-Server/} and used the GIPSY
software with the {\sc scanaid} tool to reconstruct a cut through the emission
on the exact position. This is preferred over using the IRAS Point Source
Catalogue values and works well for isolated sources, for which often reliable
flux densities can be estimated down to a level of a few 0.01 Jy at 12 and 25
$\mu$m and $\sim$0.1 Jy at 60 $\mu$m, but where clusters contain more than one
IR-bright source the flux densities can become quite unreliable and should
often be regarded as upper limits.

Most cluster IR objects have reliable MSX band A (8.3 $\mu$m) flux densities
in the MSX Point Source Catalogue (Version 2.3). However, the flux densities
for NGC\,419-IR1 and IR2, NGC\,1903-IR1 and IR2, NGC\,1978-IR3 and IR4,
HS\,327-E-IR2, SL\,519-IR2 and NGC\,2121-IR1 were estimated by differential
aperture photometry on the MSX images with respect to nearby moderately-bright
sources listed in the MSX Point Source Catalogue. The SMC cluster objects
NGC\,419-IR1 and IR2 were also measured on the original MSX images. Only a few
objects are bright enough to have reliable MSX measurements in bands C (12.1
$\mu$m), D (14.7 $\mu$m) or E (21.3 $\mu$m).

Three of the clusters, BSDL\,1837, HS\,327 and SL\,519 were covered by the ISO
mini-survey of the Magellanic Clouds (Loup et al., in preparation), performed
with the ISOCAM instrument and filters LW1 (4.5 $\mu$m) and LW10 (12.0
$\mu$m). SL\,519 is located in the prime calibration field for the survey,
which is also one of the calibration fields for the ISOGAL survey (Omont et
al.\ 2003), and was therefore also observed twice with the LW2 (6.7 $\mu$m)
filter, a second time with the LW10 filter and once with the LW3 (14.3 $\mu$m)
filter. Flux densities were measured through differential aperture photometry
with respect to moderately-bright objects in the ISO mini-survey and/or ISOGAL
Point Source Catalogues. Time-averaged images were created for SL\,519 by
rotate-shift-adding the images taken at the different epochs.

\section{Results}

%
%
\begin{table*}
\caption[]{List of cluster IR objects (J2000 coordinates, based on 2MASS), and
identifications with mid-IR sources. Mid-IR flux densities (in Jy) were
obtained from the MSX Point Source Catalogue or from the original MSX images,
and from IRAS scans; the bands are listed by their central wavelengths in
$\mu$m. Values marked with a colon are suspect. The references are as follows:
 1=Egan, Van Dyk \& Price (2001);
 2=Elias, Frogel \& Humphreys (1985);
 3=Ferraro et al.\ (1995);
 4=Frogel et al.\ (1990);
 5=Israel \& Koornneef (1991);
 6=Leisy et al.\ (1997);
 7=Lloyd-Evans (1980);
 8=Loup et al.\ (1997);
 9=Marshall et al.\ (2004);
10=Matsuura et al.\ (2002);
11=Morgan (1984);
12=Nishida et al.\ (2000);
13=Reid, Tinney \& Mould (1990);
14=Sanduleak, MacConnell \& Philip (1978);
15=Tanab\'{e} et al.\ (1997);
16=Tanab\'{e} et al.\ (1999);
17=Tanab\'{e} et al.\ (2004);
18=Trams et al.\ (1999a);
19=Trams et al.\ (1999b);
20=van Loon, Zijlstra \& Groenewegen (1999a);
21=van Loon et al.\ (1997);
22=van Loon et al.\ (1998);
23=van Loon et al.\ (1999b);
24=van Loon et al.\ (2001a);
25=van Loon et al.\ (2001b);
26=van Loon et al.\ (2003);
27=van Loon et al.\ (2005);
28=van Loon et al., in preparation;
29=Villaver, Stanghellini \& Shaw (2003);
30=Westerlund, Olander \& Hedin (1981);
31=Whitelock et al.\ (2003);
32=Wood \& Cohen (2001);
33=Wood et al.\ (1992).}
\begin{tabular}{lccccrrrrrl}
\hline\hline
\multicolumn{2}{c}{Name}               &
RA                                     &
Dec                                    &
\multicolumn{2}{c}{Identification(s)}  &
MSX                                    &
IRAS                                   &
...                                    &
...                                    &
References                             \\
Cluster                                &
\llap{IR}\#                            &
($^{\rm h}$ $^{\rm m}$ $^{\rm s}$)     &
($^\circ$ $^\prime$ $^{\prime\prime}$) &
IRAS                                   &
MSX                                    &
8.3                                    &
12                                     &
25                                     &
60                                     &
                                       \\
\hline
\multicolumn{11}{l}{\it Small Magellanic Cloud} \\
NGC\,419            &
1                   &
1 08 12.97          &
$-$72 52 44.0       &
                    &
                    &
0.058               &
\llap{$<$}0.10      &
\llap{$<$}0.14      &
                    &
12,15,17            \\
...                 &
2                   &
1 08 17.52          &
$-$72 53 09.2       &
LI-SMC\,182         &
                    &
0.183               &
0.24                &
0.19                &
$<$0.2              &
16,17               \\
\multicolumn{11}{l}{\it Large Magellanic Cloud} \\
HS\,33              &
1                   &
4 49 18.49          &
$-$69 53 14.5       &
04496$-$6958        &
1130                &
0.404               &
0.31                &
0.22\rlap{$^{\rm a}$} &
$<$0.1              &
18,19,20,22,23,31   \\
KMHK\,29\rlap{2}    &
1                   &
4 55 34.85          &
$-$69 26 55.7       &
04559$-$6931        &
1329                &
0.298               &
0.32                &
0.18                &
$<$0.2              &
30                  \\
...                 &
2                   &
4 55 41.83          &
$-$69 26 24.3       &
                    &
                    &
0.171               &
0.20\rlap{:}        &
0.08\rlap{:}        &
                    &
19,23,30            \\
KMHK\,28\rlap{5}    &
1                   &
4 55 38.98          &
$-$67 49 10.7       &
04557$-$6753        &
1238                &
0.157               &
0.24                &
0.15                &
$<$0.3              &
10,19,22,31         \\
NGC\,1783           &
1                   &
4 59 01.11          &
$-$65 58 30.3       &
                    &
1273                &
0.116               &
0.05                &
0.03                &
                    &
12,15,16,17         \\
NGC\,1852           &
1                   &
5 09 20.22          &
$-$67 47 25.0       &
05094$-$6751        &
66                  &
0.065               &
0.13                &
0.45                &
$<$0.5              &
6,11,13,14,29       \\
NGC\,1903           &
1                   &
5 17 16.33          &
$-$69 20 29.8       &
                    &
                    &
0.104               &
$<$0.20             &
$<$0.20             &
                    &
28                  \\
...                 &
2                   &
5 17 17.38          &
$-$69 20 54.6       &
                    &
                    &
0.052               &
$<$0.20             &
$<$0.20             &
                    &
28                  \\
...                 &
3                   &
5 17 22.62          &
$-$69 20 15.5       &
05176$-$6922        &
344                 &
0.301               &
0.22\rlap{:}        &
0.09\rlap{:}        &
                    &
                    \\
HS\,270             &
1                   &
5 23 53.93          &
$-$71 34 43.9       &
05246$-$7137        &
423                 &
0.161               &
0.19                &
0.40                &
2.0                 &
                    \\
SL\,453             &
1                   &
5 25 03.26          &
$-$69 26 17.0       &
                    &
484                 &
0.156               &
$<$0.10             &
$<$0.10             &
                    &
28                  \\
SL\,482             &
1                   &
5 27 17.84          &
$-$66 22 05.6       &
05273$-$6624        &
562                 &
0.233               &
$<$0.10             &
$<$0.12             &
                    &
8,13                \\
NGC\,1984           &
1                   &
5 27 40.83          &
$-$69 08 05.4       &
05280$-$6910        &
                    &
$\lsim$1.370        &
3.90                &
23.50               &
12.6                &
9,25,28,32,33       \\
...                 &
2                   &
5 27 40.11          &
$-$69 08 04.5       &
                    &
                    &
$\ll$1.370          &
                    &
                    &
                    &
2,28,30             \\
...                 &
3                   &
5 27 35.67          &
$-$69 08 56.3       &
                    &
                    &
0.157               &
0.10\rlap{:}        &
$<$1.00             &
                    &
5,6,11,14           \\
BSDL\,183\rlap{7}   &
1                   &
5 27 47.48          &
$-$69 13 20.5       &
05281$-$6915        &
588                 &
0.320               &
0.30                &
0.20                &
                    &
30                  \\
NGC\,1994           &
1                   &
5 28 21.98          &
$-$69 08 33.7       &
05287$-$6910        &
583                 &
$\lsim$0.962        &
1.07                &
1.40                &
$<$1.0              &
                    \\
NGC\,1978           &
1                   &
5 28 40.17          &
$-$66 13 54.2       &
                    &
                    &
0.077               &
0.07\rlap{:}        &
\llap{$<$}0.05      &
                    &
4,12,17,28          \\
...                 &
2                   &
5 28 47.20          &
$-$66 14 13.6       &
05287$-$6616        &
                    &
0.124               &
0.08\rlap{:}        &
$<$0.10             &
                    &
3,16,17,28          \\
...                 &
3                   &
5 29 02.41          &
$-$66 15 27.8       &
05289$-$6617        &
                    &
0.080               &
0.16                &
0.38                &
0.3\rlap{:}         &
13,19,23,27         \\
...                 &
4                   &
5 28 44.50          &
$-$66 14 04.0       &
                    &
                    &
0.070\rlap{:}       &
0.10\rlap{:}        &
0.10\rlap{:}        &
                    &
4,7,17              \\
HS\,327-E           &
1                   &
5 29 24.60          &
$-$69 55 13.4       &
05298$-$6957        &
653                 &
0.292               &
0.85\rlap{$^{\rm b}$} &
1.38\rlap{$^{\rm b}$} &
$<$3.0              &
9,19,22,23,24,25,3\rlap{3} \\
...                 &
2                   &
5 29 25.46          &
$-$69 54 52.3       &
                    &
                    &
0.030\rlap{:}       &
$\ll$0.85           &
                    &
                    &
24                  \\
SL\,519             &
1                   &
5 30 04.87          &
$-$69 56 45.2       &
                    &
654                 &
0.152               &
0.20\rlap{:}        &
0.10\rlap{:}        &
                    &
28                  \\
...                 &
2                   &
5 30 02.15          &
$-$69 56 17.2       &
                    &
                    &
$\lsim$0.030        &
0.15\rlap{:}        &
0.10\rlap{:}        &
                    &
                    \\
NGC\,2100           &
1                   &
5 42 11.57          &
$-$69 12 48.8       &
05425$-$6914        &
1435                &
0.167               &
0.30                &
0.22                &
                    &
1                   \\
NGC\,2121           &
1                   &
5 48 16.81          &
$-$71 28 39.3       &
                    &
                    &
0.028\rlap{:}       &
0.02\rlap{:}        &
$<$0.05             &
                    &
7,4,17              \\
KMHK\,16\rlap{03}   &
1                   &
6 02 31.06          &
$-$67 12 47.0       &
06025$-$6712        &
1652                &
0.251               &
0.39                &
0.28                &
$\lsim$0.2          &
21,26               \\
\hline
\end{tabular}
Notes: $^{\rm a}$ ISOPHOT $F_{25}=0.126$ Jy (Trams et al.\ 1999b); $^{\rm b}$
ISOPHOT $F_{12}=0.303$ Jy and $F_{25}=0.359$ Jy (Trams et al.\ 1999b).
\end{table*}

%
%
\begin{table*}
\caption[]{Near-IR photometry of cluster IR objects: JHK$_{\rm s}$ magnitudes
from 2MASS and/or JK$_{\rm s}$L$^\prime$ magnitudes from our own imaging
(``ESO'', and ``acq'' for the spectroscopy acquisition).}
\begin{tabular}{lcrrrrrrr}
\hline\hline
Cluster                   &
\llap{IR}\#               &
$J_{\rm 2MASS}$           &
$H_{\rm 2MASS}$           &
$K_{\rm 2MASS}$           &
$J_{\rm ESO}$             &
$K_{\rm s, ESO}$          &
$L^\prime_{\rm ESO}$      &
$L^\prime_{\rm ESO, acq}$ \\
\hline
\multicolumn{9}{l}{\it Small Magellanic Cloud} \\
NGC\,419              &
1                     &
$13.48\pm0.05$        &
$12.04\pm0.04$        &
$10.89\pm0.03$        &
$13.68\pm0.01$        &
$10.75\pm0.01$        &
$8.84\pm0.01$         &
                      \\
...                   &
2                     &
                      &
                      &
                      &
$>18.65$              &
$15.63\pm0.05$        &
$\llap{1}0.71\pm0.01$ &
                      \\
\multicolumn{9}{l}{\it Large Magellanic Cloud} \\
HS\,33                &
1                     &
$12.66\pm0.03$        &
$10.85\pm0.03$        &
$9.43\pm0.02$         &
$11.71\pm0.01$        &
$8.90\pm0.01$         &
$7.07\pm0.01$         &
                      \\
KMHK\,292             &
1                     &
$8.23\pm0.02$         &
$7.45\pm0.03$         &
$7.11\pm0.02$         &
                      &
                      &
$6.74\pm0.01$         &
                      \\
...                   &
2                     &
$8.80\pm0.03$         &
$8.03\pm0.03$         &
$7.70\pm0.02$         &
                      &
                      &
$7.07\pm0.01$         &
                      \\
KMHK\,285             &
1                     &
$>$16.18              &
$14.49\pm0.08$        &
$12.40\pm0.03$        &
                      &
$12.70\pm0.06$        &
$9.38\pm0.01$         &
                      \\
NGC\,1783             &
1                     &
$14.06\pm0.04$        &
$12.16\pm0.02$        &
$10.63\pm0.02$        &
                      &
                      &
                      &
                      \\
NGC\,1852             &
1                     &
$15.83\pm0.07$        &
$15.34\pm0.13$        &
$14.40\pm0.08$        &
$15.72\pm0.60$        &
$14.45\pm0.40$        &
$11.93\pm0.05$        &
                      \\
NGC\,1903             &
1                     &
$14.07\pm0.06$        &
$12.07\pm0.03$        &
$10.59\pm0.03$        &
                      &
                      &
$8.83\pm0.01$         &
$8.27\pm0.01$         \\
...                   &
2                     &
$13.52\pm0.04$        &
$11.82\pm0.03$        &
$10.53\pm0.03$        &
                      &
                      &
$8.55\pm0.01$         &
$9.11\pm0.01$         \\
...                   &
3                     &
$10.18\pm0.07$        &
$9.37\pm0.06$         &
$8.89\pm0.05$         &
                      &
                      &
$7.73\pm0.01$         &
                      \\
HS\,270               &
1                     &
$>$16.18              &
$15.59\pm0.16$        &
$12.98\pm0.04$        &
                      &
                      &
                      &
                      \\
SL\,453               &
1                     &
$>$15.16              &
$>$14.65              &
$14.48\pm0.10$        &
                      &
                      &
                      &
$9.43\pm0.01$         \\
SL\,482               &
1                     &
$12.42\pm0.04$        &
$11.90\pm0.04$        &
$10.80\pm0.03$        &
                      &
                      &
$8.51\pm0.01$         &
                      \\
NGC\,1984             &
1                     &
                      &
                      &
                      &
$14.45\pm0.08$        &
$12.87\pm0.02$        &
$9.44\pm0.01$         &
$9.48\pm0.01$         \\
...                   &
2                     &
$9.13\pm0.03$         &
$8.49\pm0.04$         &
$8.16\pm0.02$         &
$9.33\pm0.01$         &
$8.38\pm0.01$         &
$7.54\pm0.01$         &
$7.65\pm0.01$         \\
...                   &
3                     &
$15.06\pm0.05$        &
$14.47\pm0.06$        &
$12.89\pm0.03$        &
$15.01\pm0.01$        &
$12.83\pm0.01$        &
$9.49\pm0.02$         &
                      \\
BSDL\,1837            &
1                     &
$8.82\pm0.02$         &
$8.00\pm0.04$         &
$7.60\pm0.03$         &
                      &
                      &
$7.17\pm0.01$         &
                      \\
NGC\,1994             &
1                     &
$11.02\pm0.10$        &
$8.80\pm0.12$         &
$8.34\pm0.07$         &
                      &
                      &
$7.40\pm0.01$         &
                      \\
NGC\,1978             &
1                     &
$>$13.75              &
$13.11\pm0.07$        &
$11.73\pm0.04$        &
$14.49\pm0.01$        &
$11.05\pm0.01$        &
$8.43\pm0.01$         &
$9.08\pm0.01$         \\
...                   &
2                     &
$>$13.56              &
$14.98\pm0.14$        &
$>$12.38              &
$16.67\pm0.05$        &
$12.45\pm0.01$        &
$8.98\pm0.01$         &
$8.89\pm0.29$         \\
...                   &
3                     &
$14.77\pm0.07$        &
$13.62\pm0.07$        &
$13.02\pm0.04$        &
                      &
                      &
$11.67\pm0.03$        &
$11.56\pm0.03$        \\
...                   &
4                     &
$11.49\pm0.07$        &
$10.34\pm0.09$        &
$9.68\pm0.04$         &
$11.74\pm0.01$        &
$9.73\pm0.01$         &
$8.69\pm0.01$         &
$8.53\pm0.01$         \\
HS\,327-E             &
1                     &
$>$13.86              &
$>$12.99              &
$11.38\pm0.03$        &
                      &
$10.81\pm0.02$\rlap{$^{\rm a}$} &
$^{\rm b}$            &
                      \\
...                   &
2                     &
$11.96\pm0.03$        &
$10.95\pm0.03$        &
$10.45\pm0.02$        &
                      &
$10.60\pm0.02$\rlap{$^{\rm a}$} &
                      &
                      \\
SL\,519               &
1                     &
$>$14.43              &
$>$13.83              &
$13.10\pm0.06$        &
                      &
                      &
                      &
$9.81\pm0.01$         \\
...                   &
2                     &
$15.64\pm0.09$        &
$15.49\pm0.15$        &
$14.26\pm0.09$        &
                      &
                      &
                      &
$12.17\pm0.12$        \\
NGC\,2100             &
1                     &
$9.56\pm0.03$         &
$8.62\pm0.04$         &
$8.26\pm0.02$         &
                      &
                      &
$7.81\pm0.01$         &
                      \\
NGC\,2121             &
1                     &
$12.97\pm0.03$        &
$11.44\pm0.03$        &
$10.36\pm0.02$        &
$12.31\pm0.01$        &
$10.04\pm0.01$        &
$8.55\pm0.01$         &
                      \\
KMHK\,1603            &
1                     &
$>$17.85              &
$15.61\pm0.16$        &
$12.98\pm0.04$        &
                      &
                      &
$7.72\pm0.01$         &
$9.27\pm0.25$         \\
\hline
\end{tabular}
Notes: $^{\rm a}$ earlier observation with SOFI at ESO (van Loon et al.\
2001a); $^{\rm b}$ SAAO $L=8.55\pm0.10$ mag (Trams et al.\ 1999b).
\end{table*}

\subsection{Identification of IR objects}

The cluster IR objects and their identification with known mid-IR sources, IR
photometry and classification, and the (expected) availability of 3--4 $\mu$m
spectroscopy and/or Spitzer Space Telescope IRS observations are listed in
Tables 2--7. Table 2 includes literature references for the individual IR
objects, on a few of which we comment here:
\begin{itemize}
\item{The variable M1.5 supergiant HV\,12501 (WOH\,S\,78) has been held
responsible for the mid-IR emission in KMHK\,292 (Trams et al.\ 1999b).
However, another supergiant in the cluster, WOH\,S\,76 is brighter than
HV\,12501 both in the near-IR and in MSX band A.}
\item{NGC\,1852-IR1 is identified with the low-excitation PN SMP\,LMC\,31
(Morgan 1984).}
\item{NGC\,1984 is associated with the bright IR source IRAS\,05280$-$6910 as
well as a source of OH (Wood et al.\ 1992) and H$_2$O (van Loon et al.\ 2001b)
maser emission. ATCA observations (mentioned by Wood et al.\ 1992, but
unpublished) locate the OH maser with RA$=5^{\rm h}27^{\rm m}39.87^{\rm s}$
and Dec$=-69^\circ08^\prime06.8^{\prime\prime}$ (J2000) very near the M1 red
giant WOH\,G\,347. However, our L$^\prime$-band images reveal a very red
object at only a few arcseconds from WOH\,G\,347. Our mid-IR image taken with
TIMMI2 at a wavelength of $\lambda=11.9$ $\mu$m clearly identifies this red
star with the IRAS source: WOH\,G\,347 contributes only 5 per cent to the
total emission at this wavelength. The IRAS source and the cluster itself have
been mis-identified in the past (e.g, in Simbad) with SMP\,LMC\,64, a very low
excitation PN almost an arcmin away.}
\item{SL\,519-IR1 dominates the IR emission seen in the ISOCAM images (Fig.\
3). Another very faint, somewhat red object is visible too: SL\,519-IR2. There
are three other not very red mid-IR sources which however are each blends of
at least 2--3 bright individual stars.}
\end{itemize}

The cumulative near-IR colour-(absolute)magnitude diagrams (Fig.\ 4) of all
L$^\prime$-band detected stars in and around all observed clusters in the SMC
(open symbols) and LMC (dots) are quite similar for both Magellanic Clouds.
They show the prominent branches of RGB (up to $M_{\rm Ks}\simeq-6$ mag), AGB
(up to $M_{\rm Ks}\simeq-9$ mag), massive main-sequence stars and RSGs (the
latter dominating the IR light, at $M_{\rm Ks}\lsim-10$ mag). In the
J--K$_{\rm s}$ colour, a separate branch of massive hot stars stands out,
which is however difficult to distinguish in the K$_{\rm s}$--L$^\prime$
colours. At first glance it seems that the SMC supergiants are bluer than
similar LMC supergiants by about 0.2 mag in both J--K$_{\rm s}$ and K$_{\rm
s}$--L$^\prime$, yet the upper AGB sequences seem to coincide in the SMC and
LMC.

The bright ($M_{\rm Ks}<-10$ mag) stars with negative K$_{\rm s}$--L$^\prime$
colours are all from the clusters NGC\,1994 and NGC\,2004 for which only 2MASS
K$_{\rm s}$-band magnitudes are available: crowding is a severe problem for
these compact clusters, and several of the brightest cluster members become
blended in the 2MASS data leading to overestimates in the individual stars'
K$_{\rm s}$-band magnitudes. The L$^\prime$-band data do not suffer from the
crowding and hence the K$_{\rm s}$--L$^\prime$ colours are being
under-estimated. Because the blending of similar stars does not affect their
2MASS colours the effect is not apparent in the J--K$_{\rm s}$ colours.

Stars with colours $(J-K_{\rm s})\gsim2$ or $(K_{\rm s}-L^\prime)\gsim1$ mag
are considered to be "IR objects". These have been reddened compared to normal
photospheric colours by circumstellar or interstellar selective extinction
(notably affecting the J--K$_{\rm s}$ colour) or by circumstellar emission
giving rise to excess emission at thermal IR wavelengths (e.g., L$^\prime$).
It is worth pointing out that the reddest SMC object, NGC\,419-IR1 is as red
as the reddest LMC objects. All objects with $(K_{\rm s}-L^\prime)>3$ mag and
$M_{\rm Ks}>-6$ mag are undetected in the J-band. This includes four out of
five objects that form a clump in the colour-magnitude diagram around $(K_{\rm
s}-L^\prime)\simeq 3.4$, $M_{\rm Ks}\simeq-5.6$ mag. A few objects are
included in our analysis that are not very red but dominate the cluster's
light at mid-IR wavelengths.

%
%
\begin{table}
\caption[]{Near-IR photometry of cluster IR objects from the DENIS catalogue
of the Magellanic Clouds (Cioni et al.\ 2000).}
\begin{tabular}{lcrrr}
\hline\hline
Cluster            &
\llap{IR}\#        &
$I_{\rm DENIS}$    &
$J_{\rm DENIS}$    &
$K_{\rm s, DENIS}$ \\
\hline
HS\,33     &
1          &
14.26      &
11.94      &
8.79       \\
KMHK\,292  &
1          &
10.05      &
8.42       &
7.14       \\
...        &
2          &
10.08      &
8.62       &
7.51       \\
KMHK\,285  &
1          &
           &
15.92      &
11.28      \\
NGC\,1852  &
1          &
16.53      &
15.32      &
           \\
NGC\,1903  &
2          &
15.17      &
13.12      &
10.32      \\
...        &
3          &
10.92      &
9.78       &
8.55       \\
SL\,482    &
1          &
12.50      &
12.20      &
10.72      \\
NGC\,1984  &
2          &
10.42      &
           &
8.20       \\
...        &
3          &
15.66      &
14.61      &
           \\
BSDL\,1837 &
1          &
10.61      &
8.84       &
7.51       \\
NGC\,1978  &
3          &
15.99      &
14.32      &
12.77      \\
SL\,519    &
2          &
15.72      &
15.81      &
           \\
\hline
\end{tabular}
\end{table}

%
%
\begin{table}
\caption[]{New mid-IR photometry of cluster objects, obtained at a wavelength
of 10.6 or 11.9 $\mu$m with TIMMI2 at the ESO 3.6m telescope: flux densities
are in Jy; the bands are listed by their central wavelengths in $\mu$m. In
addition, photometry in the MSX bands C, D and E are given where available.}
\begin{tabular}{lcccccc}
\hline\hline
Cluster     &
\llap{IR}\# &
10.6        &
11.9        &
12.1        &
14.7        &
21.3        \\
\hline
NGC\,1984     &
1             &
              &
6.8$\pm$0.7   &
5.28          &
9.9           &
19.8          \\
...           &
2             &
              &
\llap{0}.39$\pm$0.0\rlap{8} &
              &
              &
              \\
BSDL\,1837    &
1             &
              &
              &
0.37\rlap{:}  &
              &
              \\
NGC\,1994     &
1             &
              &
\llap{1}.59$\pm$0.2\rlap{1} &
1.04          &
0.9           &
1.1           \\
HS\,327-E     &
1             &
              &
              &
0.44          &
              &
              \\
KMHK\,160\rlap{3} &
1             &
\llap{0}.54$\pm$0.0\rlap{6} &
              &
              &
              &
              \\
\hline
\end{tabular}
\end{table}

%
%
\begin{table}
\caption[]{Mid-IR photometry of cluster objects from ISOCAM: flux densities
are in mJy; bands are listed by the central wavelength in $\mu$m.
References are: 1=Tanab\'{e} et al.\ (2004); 2=Trams et al.\
(1999b); 3=van Loon et al.\ (2001a); 4=this work.}
\begin{tabular}{lcrrrrc}
\hline\hline
Cluster     &
\llap{IR}\# &
4.5         &
6.7         &
12.0        &
14.3        &
Reference   \\
\hline
\multicolumn{7}{l}{\it Small Magellanic Cloud} \\
NGC\,419   &
1          &
71         &
49         &
37         &
           &
1          \\
...        &
2          &
29         &
56         &
103        &
           &
1          \\
\multicolumn{7}{l}{\it Large Magellanic Cloud} \\
HS\,33     &
1          &
           &
           &
269        &
           &
2          \\
NGC\,1783  &
1          &
95         &
75         &
55         &
           &
1          \\
BSDL\,1837 &
1          &
212        &
           &
236        &
           &
4          \\
NGC\,1978  &
1          &
59         &
50         &
36         &
           &
1          \\
...        &
2          &
61         &
71         &
71         &
           &
1          \\
...        &
4          &
40         &
           &
15         &
           &
1          \\
HS\,327-E  &
1          &
187        &
544        &
433        &
698        &
3          \\
...        &
2          &
15         &
9\rlap{.5} &
5\rlap{.7} &
6\rlap{.6} &
3          \\
SL\,519    &
1          &
118        &
138        &
155        &
135        &
4          \\
...        &
2          &
1\rlap{.0} &
4\rlap{.6} &
2\rlap{.6} &
1\rlap{.9} &
4          \\
NGC\,2121  &
1          &
51         &
34         &
17         &
           &
1          \\
\hline
\end{tabular}
\end{table}

%
%
\begin{figure}[]
\centerline{\psfig{figure=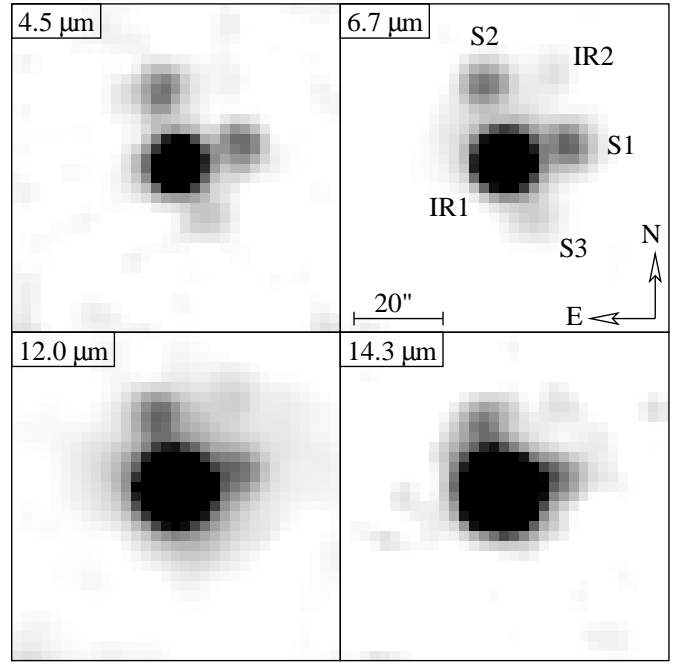,width=88mm}}
\caption[]{Mid-IR images of SL\,519 at 4.5, 6.7, 12.0 and 14.3 $\mu$m,
obtained with ISOCAM.}
\end{figure}

\subsection{IR colour-magnitude diagrams of selected clusters}

The best example in our sample of an old ($>$10 Gyr), genuine globular cluster
is the very metal-poor cluster NGC\,2210 in the outskirts of the LMC (Fig.\
5). It shows a richly populated RGB sequence, terminating abruptly at $M_{\rm
Ks}=-5.95\pm0.05$ mag. There is no sign whatsoever of an AGB sequence, nor
does it contain any IR objects.

The populous intermediate-age ($\sim$0.1--10 Gyr) clusters in our sample show
an AGB sequence which generally continues up to $M_{\rm Ks}\sim-8$ mag and
$(J-K_{\rm s})\simeq1.8$ mag (Fig.\ 6). As mentioned earlier, the AGB
sequences in the SMC clusters do not appear to be much bluer than those in the
LMC. A good example is the SMC cluster NGC\,419 of which the AGB is well
populated up to $(J-K_{\rm s})=1.9$ mag, which is redder than some LMC
clusters. NGC\,1978 is a good example where the RGB can be distinguished from
the AGB: the stellar density is high up to $M_{\rm Ks}=-6$ mag but then drops
sharply as the RGB terminates and only the AGB continues: the AGB evolution is
much more rapid than the RGB phase. Compared to NGC\,2210 the RGB of more
metal-rich intermediate-age clusters is redder in J--K$_{\rm s}$.

Many of the intermediate-age clusters contain one or more IR carbon stars
(Fig.\ 6). The most extremely dust-enshrouded IR objects are much fainter in
the K$_{\rm s}$-band than optically bright tip-AGB stars, as their dust
envelopes become optically thick even at 2 $\mu$m (e.g., NGC\,419-IR2 at
$M_{\rm Ks}=-3.4$ mag). Hot objects such as post-AGB objects and PNe are also
fainter than tip-AGB stars at near-IR wavelengths, because they shine mostly
at optical and UV wavelenghts.

The 25 Myr young SMC cluster NGC\,330 displays a bright RSG branch at
$(J-K_{\rm s})=1$ mag, and massive Main Sequence stars at $(J-K_{\rm s})=0$
mag (Fig.\ 7). The older, 130 Myr LMC cluster NGC\,1866 also contains at least
three very bright red giants. These must be massive ($M_{\rm i}>4$ M$_\odot$)
AGB stars. Such oxygen-rich M giants have near-IR colours that resemble those
of red supergiants rather than the (redder) AGB carbon stars. The
circumstellar dust shells of very luminous AGB stars and supergiants become
optically thick only at much higher mass-loss rates than envelopes of lower
luminosity AGB stars (cf.\ van Loon et al.\ 1997, 1998). The dearth of
reddened stars in young clusters does not however mean that no mass-losing
giants are present. For instance, KMHK\,292-IR1 and IR2 clearly show emission
from circumstellar dust despite not being much reddened by it.

%
%
\begin{figure*}[]
\centerline{\psfig{figure=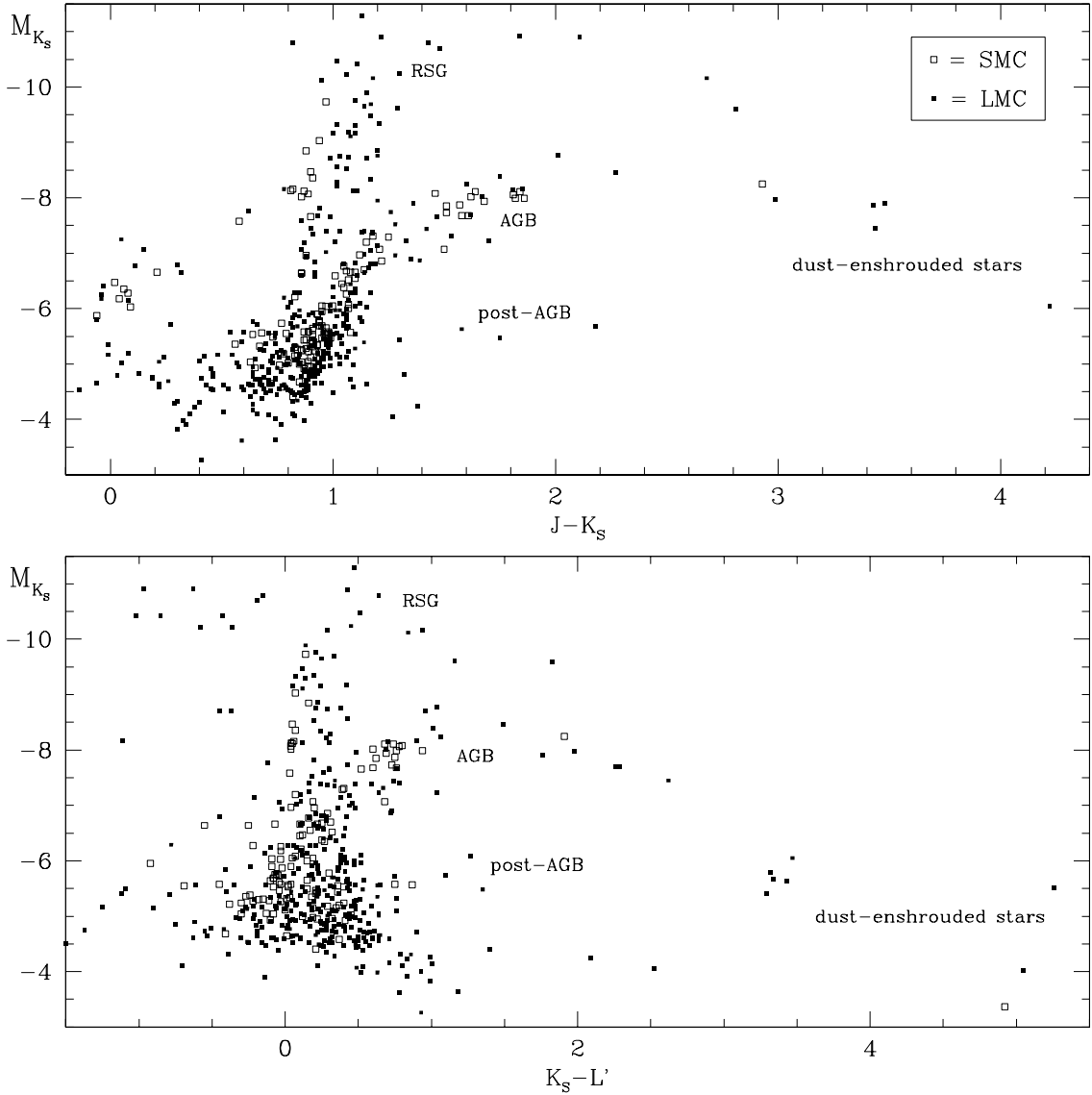,width=150mm}}
\caption[]{Cumulative colour-magnitude diagrams of absolute K$_{\rm s}$-band
magnitude versus J--K$_{\rm s}$ (top) and K$_{\rm s}$--L$^\prime$ (bottom),
for the stars detected in the L$^\prime$ band in clusters in the SMC (open
symbols) and LMC (filled symbols).}
\end{figure*}

NGC\,1903 displays two sequences, separated in J--K$_{\rm s}$ by a few tenths
of a magnitude (Fig.\ 8). The bluer of these sequences probably traces a young
population of massive AGB stars or red supergiants, whilst the redder sequence
traces an intermediate-age AGB. The brightest star, IR3 is located near the
cluster centre and seems to be associated with the bluer sequence. IR1 and IR2
are both located at the fringes of the cluster and, being carbon stars (van
Loon et al., in preparation), must be associated with the older AGB sequence.
Hence it is likely that NGC\,1903 is young and that IR1 and IR2 are associated
with an intermediate-age field population or a super-imposed cluster. This is
confirmed in the spatial distribution of the stars (Fig.\ 9): the blue Main
Sequence stars with $(J-K_{\rm s})\simeq0$ mag and the ``blue'', luminous red
giants with $(J-K_{\rm s})<1$ mag or $M_{\rm Ks}<-8$ mag form the cluster
NGC\,1903, whereas the redder, fainter red giants with $(J-K_{\rm s})>1$ mag
and $M_{\rm Ks}>-8$ mag are concentrated towards the SW of the field.

In NGC\,1984 we see a similar superposition of a young cluster and an
intermediate-age field population (Fig.\ 8). The bright red supergiant
WOH\,G\,347 (IR2) and the highly obscured object IR1 are both located near the
cluster centre. But IR3, located further out, is thought to be a PN and thus
belongs to an intermediate-age population unrelated to the cluster.

\subsection{Cluster age determinations}

%
%
\begin{figure}[]
\centerline{\psfig{figure=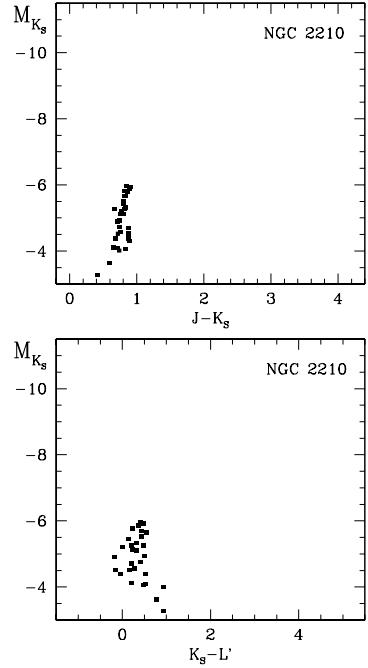,width=88mm}}
\caption[]{Example of an old cluster, NGC\,2210, displaying an RGB terminating
abruptly at $M_{\rm Ks}=-6$ mag (Section 4.2).}
\end{figure}

%
%
\begin{figure}[]
\centerline{\psfig{figure=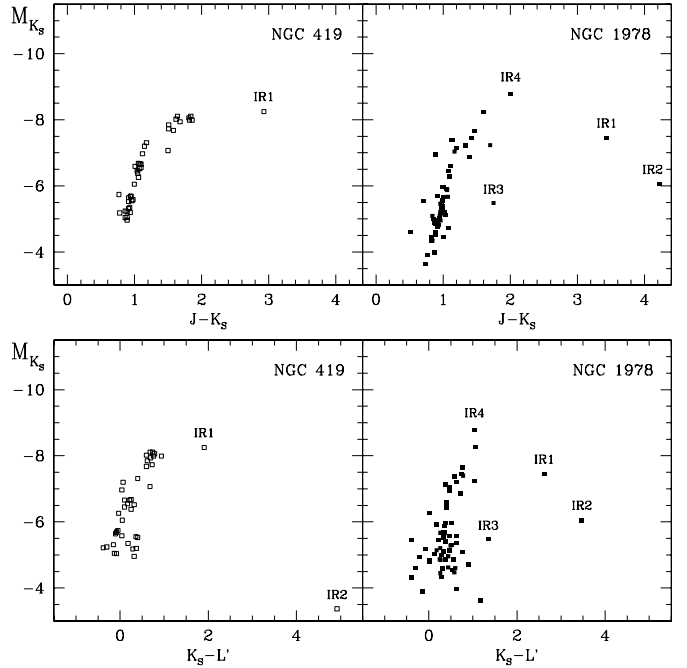,width=88mm}}
\caption[]{Examples of two populous intermediate-age clusters in the SMC and
LMC, each with several dusty stars (Section 4.2).}
\end{figure}

%
%
\begin{figure}[]
\centerline{\psfig{figure=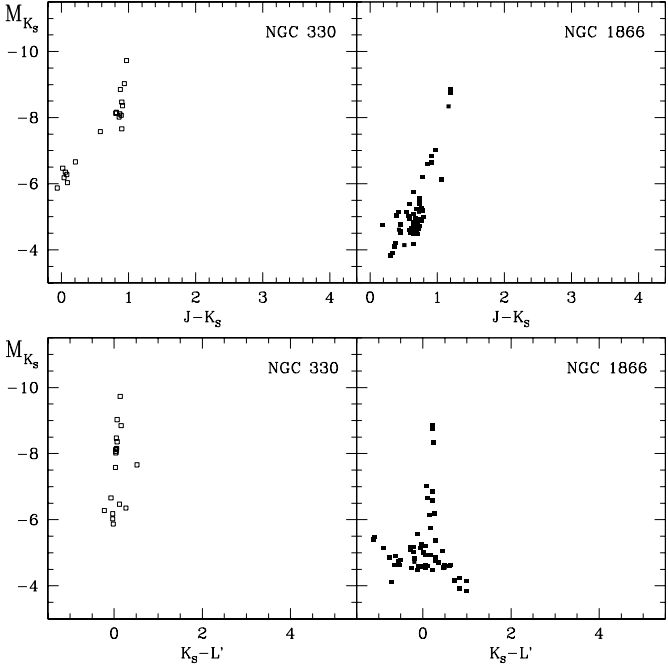,width=88mm}}
\caption[]{Examples of two populous young clusters; the SMC cluster NGC\,330
and the LMC cluster NGC\,1866 (Section 4.2).}
\end{figure}

%
%
\begin{figure}[]
\centerline{\psfig{figure=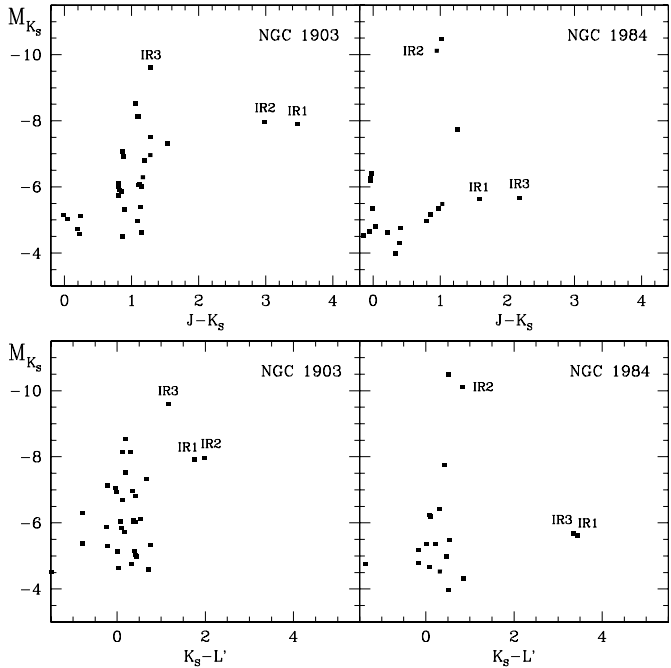,width=88mm}}
\caption[]{Examples of two LMC clusters, displaying a peculiar post-Main
Sequence morphology (Section 4.2).}
\end{figure}

%
%
\begin{figure}[]
\centerline{\psfig{figure=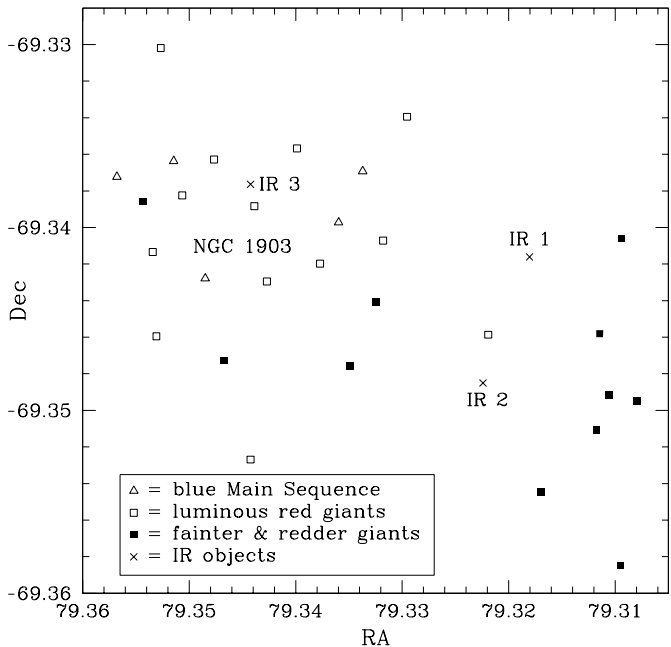,width=88mm}}
\caption[]{Spatial distribution of L$^\prime$-band detected sources near
NGC\,1903. The blue Main Sequence stars and ``blue'', luminous red giants
constitute the cluster NGC\,1903, whereas the fainter, redder red giants are
concentrated towards the SW.}
\end{figure}

%
%
\begin{figure}[]
\centerline{\psfig{figure=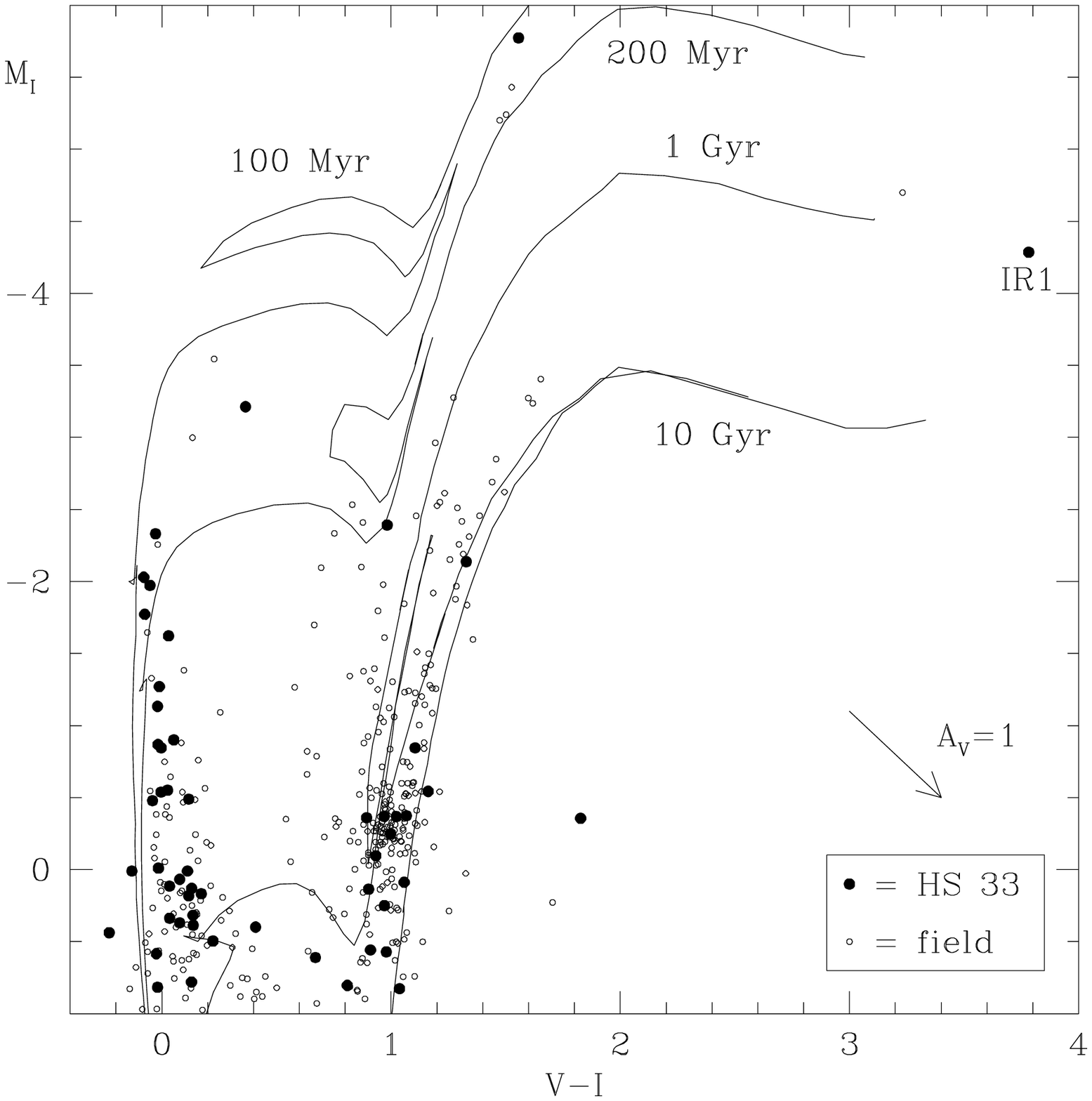,width=88mm}}
\caption[]{Optical colour-magnitude diagram for the cluster (dots) and field
(circles) around HS\,33 and IRAS\,04496$-$6958 (IR1), with overlaid isochrones
for ages of 0.1, 0.2, 1 and 10 Gyr and metallicity of [Fe/H]$=-0.4$ (Bertelli
et al.\ 1994).}
\end{figure}

For the LMC clusters HS\,33, KMHK\,292, KMHK\,285, HS\,270 and BSDL\,1837 no
age estimates could be found in the literature. We can make crude estimates of
their ages on the basis of the IR colour-magnitude diagrams, where we adopt
[Fe/H]$=-0.4$ (Table 1). BSDL\,1837 is thus found to be of intermediate age.
Because BSDL\,1837-IR1 is with $L\simeq10^5$ L$_\odot$ (see Table 8) twice as
bright as an intermediate-age AGB star would ever be, it must be much younger
than the cluster and therefore it is probably not a cluster member.

For HS\,33 we had already obtained Johnson V and Cousins I-band images with
the 0.9m Dutch telescope at ESO La Silla, Chile, on New Year's Eve 1996.
Multi-object photometry was obtained using DAOphot within MIDAS, and
calibrated against standard stars observed in the SA\,98 and T\,PHE fields.
The I versus V--I colour-magnitude diagram (Fig.\ 10) is compared with
isochrones from Bertelli et al.\ (1994) for ages of 0.1, 0.2, 1 and 10 Gyr and
a metallicity of [Fe/H]$=-0.4$. The cluster appears to have an age of
$t\sim$130$^{+70}_{-20}$ Myr, populating much more of the upper Main Sequence
around $(V-I)\sim$0 mag than of the red clump around $(V-I)\sim$1.0 and
$M_{\rm I}\sim-0.3$ mag. The $>$1 Gyr old RGB and AGB above the red clump are
almost entirely devoid of stars in the direction of HS\,33, whilst these
sequences and the red clump are heavily populated with field stars surrounding
the cluster. HS\,33-IR1 appears to lie on the red extension of the 1 Gyr
isochrone, but this is misleading as the object suffers from severe
circumstellar extinction and its colour and magnitude can be traced back to
the upper AGB of at most a few 100 Myr old. This is supporting evidence, both
for the age of HS\,33 and for IR1's association with the cluster.

%
%
\begin{figure}[]
\centerline{\psfig{figure=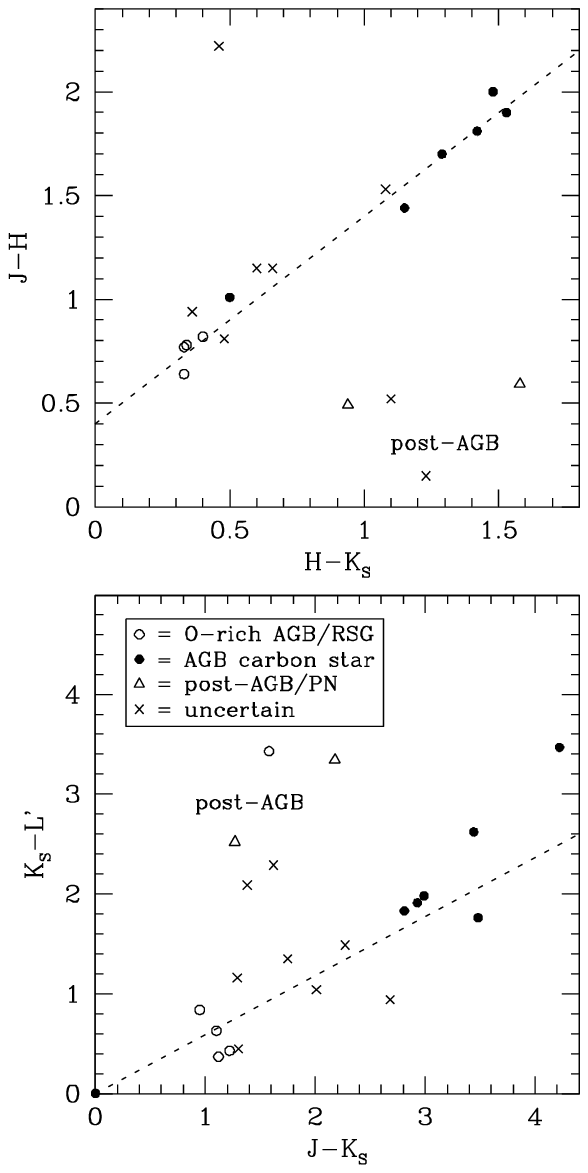,width=88mm}}
\caption[]{Near-IR colour-colour diagrams to probe the nature of the stellar
photospheres of the cluster IR objects.}
\end{figure}

\subsection{Photometric classification of cluster IR objects}

The locations of dusty stars in IR colour-magnitude and colour-colour diagrams
may clarify their nature. For instance, dust-enshrouded stars delineate a
sequence in a J--H versus H--K$_{\rm s}$ or K$_{\rm s}$--L$^\prime$ versus
J--K$_{\rm s}$ diagram (Fig.\ 11). Oxygen-rich star are found at the blue end
of this sequence and carbon stars at the red end (cf.\ Frogel \& Cohen 1982).
The confirmed PNe, NGC\,1852-IR1 and NGC\,1984-IR3 lie in a disparate region
of the near-IR colour-colour diagram, ``below'' the sequence in the J--H
versus H--K$_{\rm s}$ diagram (cf.\ Fig.\ 7 in Ferraro et al.\ 1995) and
``above'' the sequence in the K$_{\rm s}$--L$^\prime$ versus J--K$_{\rm s}$
diagram. Whitelock (1985) shows that PNe can assume negative J--H colours if
the He\,{\sc i} 1.083 $\mu$m line is included in the J-band filter. This is
the case in the SAAO filter suite but the 2MASS and ESO (SOFI) J-band filters
do not include this line.

%
%
\begin{figure*}[]
\centerline{\psfig{figure=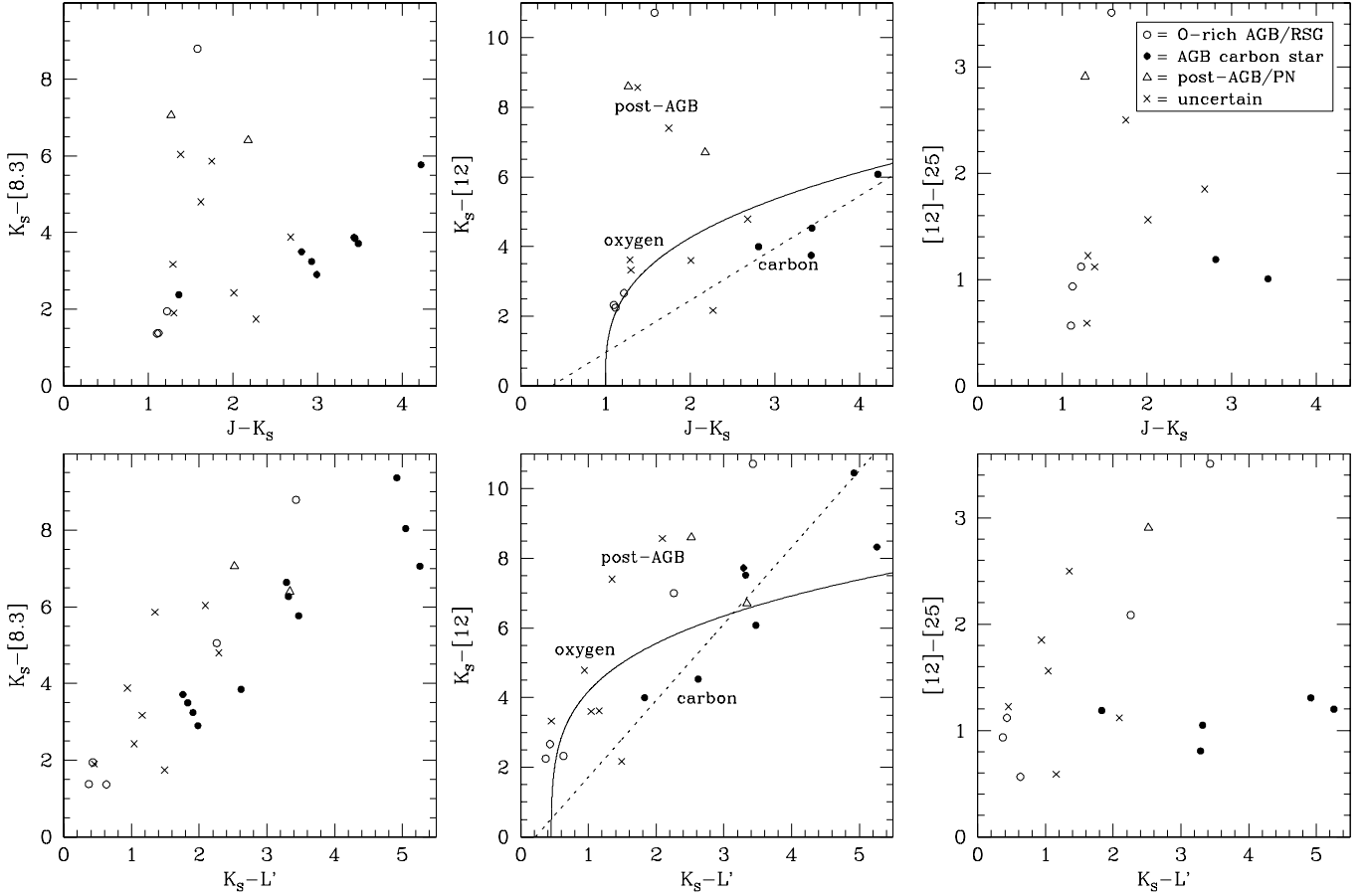,width=180mm}}
\caption[]{Near/mid-IR colour-colour diagrams that probe the nature of the
circumstellar dust shells of the cluster IR objects.}
\end{figure*}

In the K$_{\rm s}$--[12] and K$_{\rm s}$--[8.3] versus J--K$_{\rm s}$ and
versus K$_{\rm s}$--L$^\prime$ diagrams (Fig.\ 12), carbon stars and
oxygen-rich red giants follow unique sequences chiefly as a function of
optical depth. These sequences intersect each other and identification is
therefore not always conclusive. We draw the sequences for oxygen-rich AGB
stars (solid line) and carbon stars (dotted line) from van Loon et al.\ (1997)
and Trams et al.\ (1999b), after small adjustments to account for differences
in photometric systems. Post-AGB objects stand out prominently in the K$_{\rm
s}$--[12] versus J--K$_{\rm s}$ diagram (van Loon et al.\ 1997) as they have
strong dust emission but little extinction. Carbon stars also tend to occupy a
distinct region in the [12]--[25] versus J--K$_{\rm s}$ and versus K$_{\rm
s}$--L$^\prime$ diagrams (Fig.\ 12) because their circumstellar envelopes are
warmer and become optically thick more easily, and when oxygen-rich envelopes
become optically thick at mid-IR wavelengths the silicate absorption causes a
reddening of the [12]--[25] colour.

The IR diagrams suggest that NGC\,1903-IR3, NGC\,1994-IR1 and NGC\,2100-IR1
are oxygen-rich stars. HS\,270-IR1, NGC\,1984-IR1, NGC\,1978-IR3 and
SL\,482-IR1 have IR colours similar to the known low-excitation PNe
NGC\,1852-IR1 and NGC\,1984-IR3, and might have a detached dust shell.

\section{Luminosities and mass-loss rates}

\subsection{Modelling the spectral energy distributions}

The spectral energy distributions of the IR objects were modelled with the
radiative transfer code {\sc dusty} (Ivezi\'{c}, Nenkova \& Elitzur 1999). The
density distribution is based upon a hydrodynamic computation of a dust-driven
wind at constant mass-loss rate. The model was then scaled to match the
overall observed SED, knowing the distances to the SMC and LMC objects, which
then yields an accurate measurement of the bolometric luminosity.

A blackbody was used to represent the underlying stellar radiation field, with
a temperature of $T_{\rm eff}=2500$ K in the case of carbon stars and $T_{\rm
eff}=3000$ K in the case of oxygen-rich, M-type stars unless the actual
temperatures were known or the fit required a hotter star.

For almost all objects the chemistry is known, and by default we chose
amorphous carbon dust (Henning \& Mutschke 1997) or astronomical silicate
(Draine \& Lee 1984) for carbon-rich and oxygen-rich objects, respectively.
The extreme OH/IR objects NGC\,1984-IR1 and HS\,327-E-IR1 were better fit with
warm, oxygen-deficient silicate from Ossenkopf, Henning \& Mathis (1992), and
the peculiar carbon-rich object NGC\,1978-IR3 required the inclusion of 20 per
cent silicon carbide (SiC, from P\'{e}gouri\'{e} 1988). A fit to SL\,482-IR1
was obtained with cold, oxygen-rich silicate from Ossenkopf et al.\ (1992) and
a very thin shell geometry to mimic an isothermal dust envelope at $T_{\rm
dust}=750$ K. We used either a single dust grain size, or a standard MRN grain
size distribution (a power-law with exponent $-3.5$; Mathis, Rumpl \&
Nordsieck 1977).

To obtain the mass-loss rate one has to know the dust grain density, for which
we adopt $\rho_{\rm grain}=3$ g cm$^{-3}$, and the gas-to-dust mass ratio.
Very little is known about the latter, which may vary according to the
environmental conditions --- metallicity and chemistry, radiation field, and
evolutionary state. We adopt a reference value of $\psi_\odot=\rho_{\rm
gas}/\rho_{\rm dust}=200$ for solar metallicity, and scale this value
according to the cluster metallicity (Table 1) as $\psi=\psi_\odot10^{-{\rm
[Fe/H]}}$.

The input parameters and output results are summarised in Table 8. Here we
also list the initial mass for the cluster member descended from the heaviest
Main-Sequence progenitor, $M_{\rm i}$, which is taken from the Bertelli et
al.\ (1994) isochrones. These masses are representative for the cluster IR
objects that are expected to be the most evolved stars in the clusters. For
comparison we also list the Main-Sequence Turn-Off masses, defined here as the
mass of the Main Sequence star that has reached its (first) maximum in
effective temperature as its core becomes depleted in hydrogen.

%
%
\begin{table}
\caption[]{List of cluster IR objects and their classification. Objects for
which an L-band spectrum is available are referred to under ``Lspec'', where
1=van Loon et al.\ (1999a); 2=Matsuura et al.\ (2002); 3=van Loon et al.\
(2003); 4=van Loon et al.\ (in preparation). Objects which are targets for the
Spitzer Space Telescope are listed under ``SST'' by their SST programme
number, where the Principal Investigators are J.\ Houck (\#103), F.\ Kemper
(\#1094) and P.\ Wood (\#3505).}
\begin{tabular}{lclcr}
\hline\hline
Cluster                                &
\llap{IR}\#                            &
Classification                         &
Lspec                                  &
SST                                    \\
\hline
\multicolumn{4}{l}{\it Small Magellanic Cloud} \\
NGC\,419           &
1                  &
AGB carbon         &
                   &
3505               \\
...                &
2                  &
AGB carbon         &
                   &
3505               \\
\multicolumn{4}{l}{\it Large Magellanic Cloud} \\
HS\,33             &
1                  &
AGB carbon star    &
1,2                &
1094               \\
KMHK\,292          &
1                  &
M supergiant       &
                   &
                   \\
...                &
2                  &
M1.5 supergiant    &
                   &
                   \\
KMHK\,285          &
1                  &
AGB carbon star    &
2                  &
3505               \\
NGC\,1783          &
1                  &
AGB carbon star    &
                   &
$\dagger$          \\
NGC\,1852          &
1                  &
PN                 &
                   &
103                \\
NGC\,1903          &
1                  &
AGB carbon star    &
4                  &
                   \\
...                &
2                  &
AGB carbon star    &
4                  &
                   \\
...                &
3                  &
RSG? Oxygen AGB?   &
                   &
                   \\
HS\,270            &
1                  &
(post-)AGB star?   &
                   &
                   \\
SL\,453            &
1                  &
AGB carbon star    &
4                  &
                   \\
SL\,482            &
1                  &
Hot and dusty?     &
                   &
                   \\
NGC\,1984          &
1                  &
OH/IR star         &
4                  &
1094               \\
...                &
2                  &
M1 (super)giant    &
4                  &
                   \\
...                &
3                  &
PN                 &
                   &
                   \\
BSDL\,1837         &
1                  &
M1 (super)giant    &
                   &
                   \\
NGC\,1994          &
1                  &
RSG? Oxygen AGB?   &
                   &
                   \\
NGC\,1978          &
1                  &
AGB carbon star    &
4                  &
3505               \\
...                &
2                  &
AGB carbon star    &
4                  &
3505               \\
...                &
3                  &
(post-)AGB carbon \rlap{star} &
                   &
                   \\
...                &
4                  &
C? Oxygen-rich?    &
                   &
3505               \\
HS\,327-E          &
1                  &
OH/IR star         &
1                  &
                   \\
...                &
2                  &
AGB carbon star    &
                   &
                   \\
SL\,519            &
1                  &
AGB carbon star    &
4                  &
                   \\
...                &
2                  &
Post-AGB? Carbon \rlap{star?} &
                   &
                   \\
NGC\,2100          &
1                  &
RSG? Oxygen AGB?   &
                   &
                   \\
NGC\,2121          &
1                  &
AGB carbon star?   &
                   &
                   \\
KMHK\,1603         &
1                  &
AGB carbon star    &
3                  &
                   \\
\hline
\end{tabular}
$\dagger$ = SST \#3725 of P.\ Goudfrooij maps NGC\,1783 with IRAC
\end{table}

\subsection{Luminosities and nature of cluster IR objects}

Not surprisingly, all cluster IR objects are luminous (Fig.\ 13b). Where the
cluster has a RGB, the cluster IR objects were always more luminous than the
RGB tip. The only exception is the faint object SL\,519-IR2, of which the
exact nature is uncertain. Most objects straddle along the maximum luminosity
reached within a cluster (the curve in Fig.\ 13b, derived from Bertelli et
al.\ 1994).

It is especially interesting to note that the luminous carbon star HS\,33-IR1
matches very well the expected luminosity at the tip of the AGB for the
cluster HS\,33. This lends support to the interpretation of it having evolved
from a massive AGB star progenitor ($M_{\rm i}>4$ M$_\odot$).

All dust-enshrouded stars with progenitors of $1.3<M_{\rm i}<2.2$ M$_\odot$
are carbon stars, whilst stars with progenitor masses of $M_{\rm i}>6$
M$_\odot$ are always oxygen-rich.

Our sample includes four clusters with $M_{\rm i}\simeq4$ M$_\odot$. No
IR-excess objects were found in the SMC cluster NGC\,458, despite its richness
and rather high metallicity. The LMC cluster objects SL\,453-IR1,
HS\,327-E-IR2 and SL\,519-IR1 are carbon stars. These are the only cluster IR
objects that have luminosities that are significantly lower than the maximum
possible luminosity. However, the brightest IR object in these clusters,
HS\,327-E-IR1 is an oxygen-rich object of which the luminosity is right at the
tip of the AGB for that cluster age. It has been argued (van Loon et al.\
2001a) that the coincidence of carbon stars and more luminous oxygen-rich
stars in populations with an age of $t\simeq2\times10^8$ yr suggests that
$M_{\rm i}=4$ M$_\odot$ is the threshold of Hot Bottom Burning (HBB; Boothroyd
\& Sackmann 1992), with the luminous oxygen-rich AGB star descendent from a
progenitor with a slightly larger mass than this threshold and the carbon star
having a slightly lower mass. Alternatively, a second parameter might be
involved, such as rotation or a metallicity spread.

The late stages in the evolution of stars with progenitor masses in the range
$M_{\rm i}=5$--8 M$_\odot$ (shaded regime in Fig.\ 13) are uncertain, with the
Bertelli et al.\ (1994) models suggesting that the maximum progenitor mass of
an AGB star is $M_{\rm i}\simeq5$ M$_\odot$. The oxygen-rich red giant
NGC\,1903-IR3 is the only cluster IR object found in the three clusters in
this mass range. It is more luminous than expected, which may be due to HBB
(the Bertelli tracks do not include HBB).

Unfortunately we did not find cluster IR objects in our sample around $M_{\rm
i}\simeq10$ M$_\odot$. Such stars are believed to become ``super-AGB'' stars
that follow a thermal-pulsing AGB evolution but go on to ignite core-carbon
burning (Ritossa, Garcia-Berro \& Iben 1996). The supergiants of $M_{\rm
i}=14$--19 M$_\odot$ that we did find have luminosities that match very well
the expected luminosities for the clusters they are associated with.

%
%
\begin{table*}
\caption[]{Initial metallicity, [Fe/H], and mass, $M_{\rm i}$, and
Main-Sequence Turn-Off mass, $M_{\rm TO}$, and the input and results of the
modelling with {\sc dusty}: stellar effective temperature, $T_{\rm eff}$,
temperature at the inner radius of the dust envelope, $T_{\rm dust}$, dust
type (Sil-DL=astronomical silicate, Draine \& Lee 1984; Sil-Oc=cold
oxygen-rich silicate, Ossenkopf et al.\ 1992; Sil-Ow=warm oxygen-deficient
silicate, Ossenkopf et al.\ 1992; AmC=amorphous carbon, Henning \& Mutschke
1997; SiC=silicon carbide, P\'{e}gouri\'{e} 1988), grain size, $a$ (where a
range is given, these represent the minimum and maximum grain sizes for a
standard MRN distribution --- Mathis et al.\ 1977), bolometric luminosity,
$L$, and total (gas+dust) mass-loss rate, $\dot{M}$.}
\begin{tabular}{lccccccccccc}
\hline\hline
Cluster             &
\llap{IR}\#         &
[Fe/H]              &
$M_{\rm i}$         &
$M_{\rm TO}$        &
$T_{\rm eff}$       &
$T_{\rm dust}$      &
dust                &
$a$                 &
$\log(L)$           &
$\log(\dot{M})$     &
cluster             \\
                    &
                    &
                    &
(M$_\odot$)         &
(M$_\odot$)         &
(K)                 &
(K)                 &
type                &
($\mu$m)            &
(L$_\odot$)         &
(M$_\odot$/yr)      &
member              \\
\hline
\multicolumn{12}{l}{\it Small Magellanic Cloud} \\
NGC\,419            &
1                   &
$-0.6$\rlap{0}      &
1.9\rlap{8}         &
1.5\rlap{7}         &
2800                &
\llap{1}000         &
AmC                 &
0.2                 &
4.00                &
$-$5.00             &
yes                 \\
...                 &
2                   &
...                 &
...                 &
...                 &
2800                &
700                 &
AmC                 &
0.1                 &
3.85                &
$-$4.34             &
yes                 \\
\multicolumn{12}{l}{\it Large Magellanic Cloud} \\
HS\,33              &
1                   &
$-0.4$              &
4.8                 &
4.0                 &
2500                &
720                 &
AmC                 &
0.01--0.1           &
4.61                &
$-$4.54             &
yes\rlap{?}         \\
KMHK\,292           &
1                   &
$-0.4$              &
\llap{1}5.0         &
\llap{1}1.9         &
3000                &
400                 &
Sil-DL              &
0.1                 &
5.37                &
$-$4.62             &
yes                 \\
...                 &
2                   &
...                 &
...                 &
...                 &
3000                &
350                 &
Sil-DL              &
0.1                 &
5.17                &
$-$5.20             &
yes                 \\
KMHK\,285           &
1                   &
$-0.4$              &
2.2\rlap{1}         &
1.7\rlap{3}         &
2800                &
600                 &
AmC                 &
0.01--0.1           &
3.99                &
$-$4.57             &
yes                 \\
NGC\,1783           &
1                   &
$-0.7$\rlap{5}      &
1.3\rlap{2}         &
1.1\rlap{6}         &
2800                &
1100                &
AmC                 &
0.1                 &
3.93                &
$-$4.90             &
yes                 \\
NGC\,1852           &
1                   &
$-0.8$\rlap{5}      &
1.3\rlap{1}         &
1.1\rlap{6}         &
\llap{2}8600        &
500                 &
AmC                 &
10                  &
4.29                &
$-$4.25             &
yes                 \\
NGC\,1903           &
1                   &
$-0.4$              &
6.3                 &
5.3                 &
2800                &
\llap{1}300         &
AmC                 &
0.01--0.1           &
3.99                &
$-$5.11             &
no                  \\
...                 &
2                   &
...                 &
...                 &
...                 &
2800                &
\llap{1}500         &
AmC                 &
0.01--0.1           &
3.84                &
$-$5.43             &
no                  \\
...                 &
3                   &
...                 &
...                 &
...                 &
3000                &
\llap{1}200         &
Sil-DL              &
0.01--0.1           &
4.66                &
$-$4.87             &
yes                 \\
HS\,270             &
1                   &
$-0.4$              &
2.2\rlap{1}         &
1.7\rlap{3}         &
2500                &
225                 &
AmC                 &
0.01--0.3           &
3.94                &
$-$4.15             &
yes                 \\
SL\,453             &
1                   &
$-0.4$              &
4.0                 &
3.3                 &
2500                &
\llap{1}100         &
AmC                 &
0.01--0.1           &
3.70                &
$-$4.75             &
yes                 \\
SL\,482             &
1                   &
$-0.4$              &
\llap{1}3.7         &
\llap{1}1.1         &
\llap{2}0000        &
750                 &
Sil-Oc              &
0.1                 &
5.19                &
$-$4.02             &
yes                 \\
NGC\,1984           &
1                   &
$-0.9$\rlap{0}      &
\llap{1}9.1         &
\llap{1}4.3         &
\llap{1}5000        &
400                 &
Sil-Ow              &
0.01--0.1           &
5.39                &
$-$2.38             &
yes                 \\
...                 &
2                   &
...                 &
...                 &
...                 &
4000                &
\llap{1}000         &
Sil-DL              &
0.1                 &
5.10                &
$-$4.25             &
yes                 \\
...                 &
3                   &
...                 &
...                 &
...                 &
\llap{3}1000        &
650                 &
AmC                 &
0.1--0.2            &
4.47                &
$-$4.91             &
no                  \\
BSDL\,1837          &
1                   &
$-0.4$              &
2.2\rlap{1}         &
1.7\rlap{3}         &
2500                &
\llap{1}000         &
Sil-DL              &
0.1                 &
5.04                &
$-$5.03             &
no                  \\
NGC\,1994           &
1                   &
$-0.2$\rlap{4}      &
\llap{1}4.1         &
\llap{1}1.7         &
2500                &
600                 &
Sil-DL              &
0.1                 &
4.90                &
$-$4.02             &
yes                 \\
NGC\,1978           &
1                   &
$-0.6$\rlap{6}      &
1.4\rlap{2}         &
1.3\rlap{0}         &
2500                &
\llap{1}000         &
AmC                 &
0.1                 &
3.73                &
$-$5.05             &
yes                 \\
...                 &
2                   &
...                 &
...                 &
...                 &
2500                &
\llap{1}100         &
AmC                 &
0.1                 &
3.77                &
$-$4.80             &
yes                 \\
...                 &
3                   &
...                 &
...                 &
...                 &
2500                &
300                 &
AmC+SiC             &
0.01--5.0           &
3.70                &
$-$4.39             &
yes\rlap{?}         \\
...                 &
4                   &
...                 &
...                 &
...                 &
2500                &
\llap{1}000         &
AmC                 &
0.1                 &
4.19                &
$-$5.46             &
yes                 \\
HS\,327-E           &
1                   &
$-$0.4              &
4.0                 &
3.3                 &
2500                &
800                 &
Sil-Ow              &
0.01--0.1           &
4.45                &
$-$3.90             &
yes                 \\
...                 &
2                   &
...                 &
...                 &
...                 &
2500                &
200                 &
AmC                 &
0.1                 &
3.89                &
$-$5.83             &
yes                 \\
SL\,519             &
1                   &
$-0.4$              &
4.0                 &
3.3                 &
2500                &
650                 &
AmC                 &
0.01--0.1           &
3.84                &
$-$4.67             &
yes                 \\
...                 &
2                   &
...                 &
...                 &
...                 &
2500                &
830                 &
AmC                 &
0.2                 &
2.58                &
$-$6.26             &
yes\rlap{?}         \\
NGC\,2100           &
1                   &
$-0.3$\rlap{2}      &
\llap{1}4.3         &
\llap{1}1.7         &
2500                &
300                 &
Sil-DL              &
0.1                 &
4.87                &
$-$4.69             &
yes                 \\
NGC\,2121           &
1                   &
$-0.6$\rlap{1}      &
1.3\rlap{1}         &
1.1\rlap{5}         &
2500                &
\llap{1}200         &
AmC                 &
0.2                 &
3.95                &
$-$5.36             &
yes                 \\
KMHK\,1603          &
1                   &
$-0.6$\rlap{0}      &
2.1\rlap{7}         &
1.6\rlap{8}         &
2500                &
600                 &
AmC                 &
0.1                 &
4.09                &
$-$4.33             &
yes                 \\
\hline
\end{tabular}
\end{table*}

\subsection{The contribution of dust-enshrouded red giants to the integrated
cluster IR brightness}

Dust-enshrouded giants do not contribute to the optical brightness of the
cluster they are in, but they dominate the integrated cluster brightness at
mid-IR wavelengths of $\lambda\gsim8$ $\mu$m even in populous clusters. Thus,
the mid-IR brightness of a cluster cannot be reliably predicted as the rare
occurrence of a dust-enshrouded red giant is a stochastic event. However, if
the combined effect of many clusters is considered, then a sufficient number
of dust-enshrouded red giants will always be present and the fluctuations will
be small. Computations of Single Stellar Populations (Bressan, Granato \&
Silva 1998; Piovan, Tantalo \& Chiosi 2003) may therefore yield reliable
predictions for elliptical galaxies but not for star clusters.

With a central wavelength of 3.5 $\mu$m, IRAC channel 1 on-board the Spitzer
Space Telescope is similar to the L-band; what is the contribution of the
dust-enshrouded red giants to the cluster brightness in the L$^\prime$-band?
Consider the example of NGC\,419, a populous cluster with two very red
dust-enshrouded carbon stars. The combined brightness of these two IR objects
is $L^\prime_{\rm IR}=8.66$ mag. The combined brightness of all other stars in
the ISAAC fields above the RGB tip ($M_{L^\prime}=-6.4$ mag; see Section 5.5)
is $L^\prime_{>\rm RGB}=7.58$ mag. Including all fainter stars down to
$L^\prime_{\rm limit}\sim14$ mag the cluster brightness becomes $L^\prime_{\rm
cluster}=7.49$ mag. This estimate excludes the (small) contribution of even
fainter cluster members as well as stars outside of the ISAAC field of view.
Thus, the two IR objects contribute $<30$ per cent to the integrated cluster
brightness. For smaller clusters the contribution can be much larger and the
cluster L$^\prime$-band brightness becomes completely stochastic.

%
%
\begin{figure*}[]
\centerline{\psfig{figure=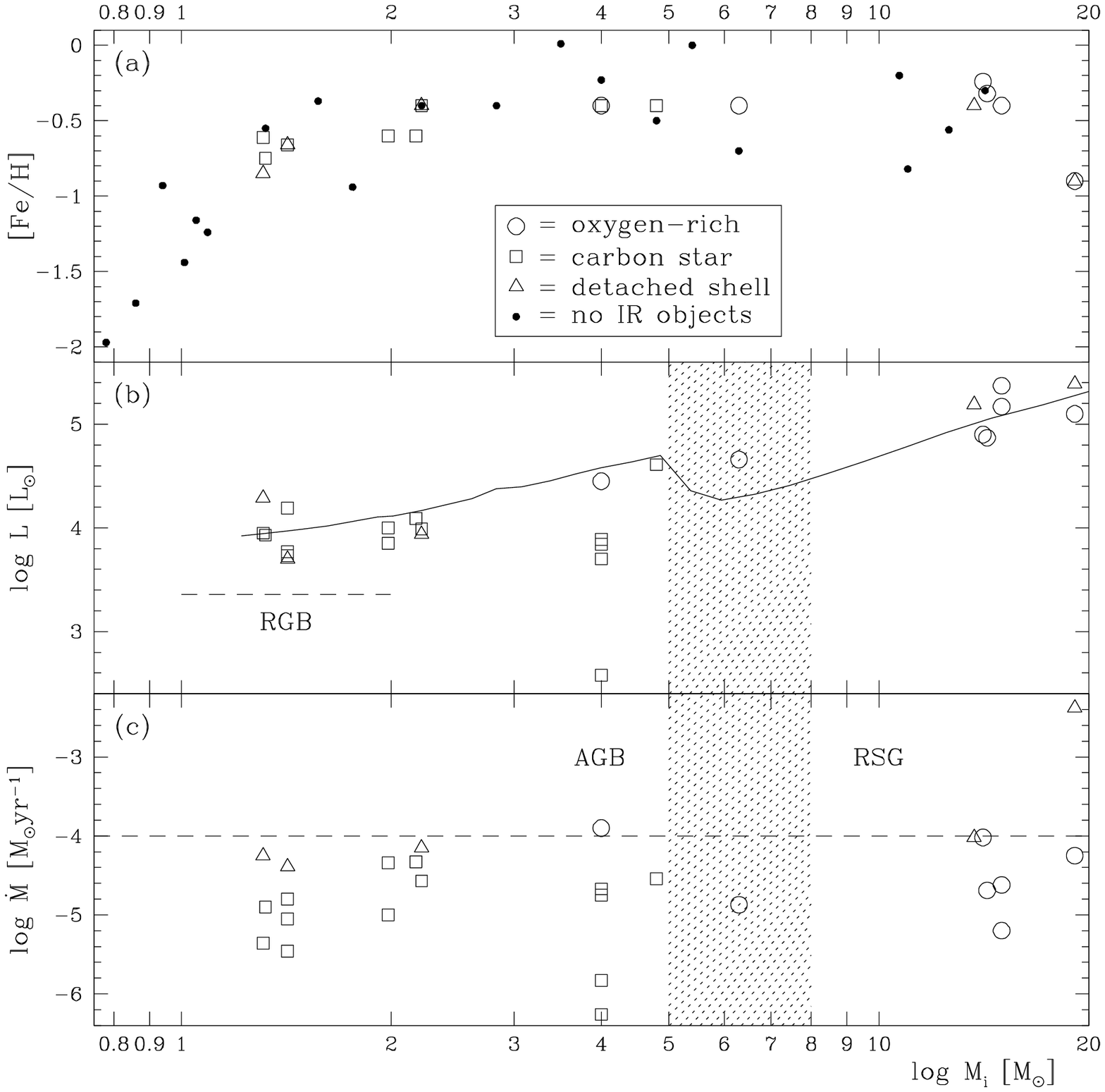,width=150mm}}
\caption[]{Luminosities and mass-loss rates of cluster IR objects as a
function of their initial mass. Panel ({\bf a}) displays all clusters in a
mass-metallicity diagram. The mass corresponds to the initial mass of the
currently most evolved star in the cluster. Oxygen-rich giants, carbon stars
and objects with detached shells are represented by discs, boxes and
triangles, respectively, where open symbols indicate that the IR object is
most likely not a physical member of the cluster. The curve in the luminosity
diagram ({\bf b}) traces the maximum luminosity achieved, from Bertelli et
al.\ (1994) models of [Fe/H]=$-0.4$ and $-0.7$ for masses larger and smaller
than 2 M$_\odot$, respectively. The shaded area indicates the regime between
$M_{\rm i}=5$ and 8 M$_\odot$ where the transition between AGB and RSG is
expected. In panel ({\bf c}) the dashed horizontal at $\dot{M}=10^{-4}$
M$_\odot$ yr$^{-1}$ is for reference only.}
\end{figure*}

\subsection{Mass-loss rates of cluster IR objects}

Most cluster IR objects are fairly evenly distributed over more than a decade
in mass-loss rate with a maximum of $\dot{M}\simeq10^{-4}$ M$_\odot$
yr$^{-1}$ (Fig.\ 13c). The spread in mass-loss rates for a given progenitor
mass is due to evolutionary effects, as the stars are not always captured at
the highest mass-loss rate. This effect can be traced by the position of the
star in the Hertzsprung-Russell diagram, and van Loon et al.\ (1999b, 2005)
indeed show empirical evidence for a dependence of the mass-loss rate on the
bolometric luminosity and stellar effective temperature.

However, upon closer inspection it seems that the lower mass progenitors,
$M_{\rm i}\simeq1.4$ M$_\odot$, reach mass-loss rates that are lower than
those of somewhat higher mass progenitors, $M_{\rm i}\simeq2$ M$_\odot$,
whilst the highest mass-loss rate occurs at an even higher progenitor mass of
$M_{\rm i}\simeq4$ M$_\odot$. A similar tendency is seen amongst the massive
progenitors of $M_{\rm i}>13$ M$_\odot$. As both the (maximum) luminosity and
cluster metallicity increase with increasing progenitor mass, it is difficult
to ascertain the physical cause for the dependence of the mass-loss rate on
progenitor mass.

Objects with detached shells (triangles in Fig.\ 13) are always the stars with
the highest mass-loss rate at a given progenitor mass. For these stars the
mass-loss rate refers to the detached dust shell and is thus a historic
account of the mass loss experienced some time in the past (typically a few
centuries ago). Hence the most likely explanation is that these are stars
which have just emerged from the most intense superwind phase in their
evolution. Amongst these, the luminous OH/IR object NGC\,1984-IR1 had an
extremely high mass-loss rate of $\dot{M}\simeq4\times10^{-3}$ M$_\odot$
yr$^{-1}$. Modelling the SED using a cool star of 2500 K hardly affects the
estimate of its luminosity but the mass-loss rate would be even higher by
$\sim$50 per cent.

%
%
\begin{table*}
\caption[]{Total cluster mass, mass-loss rate and feedback timescale, and
number of cluster IR objects, number of stars brighter in L$^\prime$ than the
RGB tip, their ratio and the cluster mass per super-RGB-tip or IR star, for
different initial mass or cluster age.}
\begin{tabular}{cccccccccc}
\hline\hline
$M_{\rm i}$                          &
$t_{\rm cluster}$                    &
$\sum M_{\rm cluster}$               &
$\sum \dot{M}$                       &
$t_{\rm mass\ loss}$                 &
$N_{> \rm RGB}$                      &
$N_{\rm IR}$                         &
\underline{$N_{\rm IR}$}             &
$\sum M_{\rm cluster}/N_{> \rm RGB}$ &
$\sum M_{\rm cluster}/N_{\rm IR}$    \\
(M$_\odot$)           &
(Gyr)                 &
(M$_\odot$)           &
(M$_\odot$ yr$^{-1}$) &
(Gyr)                 &
                      &
                      &
$N_{> \rm RGB}$       &
(M$_\odot$)           &
(M$_\odot$)           \\
\hline
$>$10              &
$<$0.03            &
$1.3\times10^5$    &
$4.5\times10^{-3}$ &
0.03               &
68                 &
7                  &
0.10               &
$1.9\times10^3$    &
$1.9\times10^4$    \\
3--1\rlap{0}       &
\llap{0}.03--0\rlap{.4} &
$2.2\times10^5$    &
$2.1\times10^{-4}$ &
1                  &
48                 &
7                  &
0.15               &
$4.6\times10^3$    &
$3.1\times10^4$    \\
\llap{1.}8--3      &
0.4--1\rlap{.4}    &
$2.4\times10^5$    &
$2.0\times10^{-4}$ &
1                  &
31                 &
5                  &
0.16               &
$7.7\times10^3$    &
$4.8\times10^4$    \\
\llap{1.}3--1\rlap{.8} &
1.4--3             &
$1.7\times10^6$    &
$1.4\times10^{-4}$ &
10                 &
36                 &
7                  &
0.19               &
$4.7\times10^4$    &
$2.4\times10^5$    \\
$<$1.3             &
$>$3               &
$1.6\times10^6$    &
$<10^{-4}$         &
\llap{$>$}10       &
5                  &
0                  &
0.00               &
$3.2\times10^5$    &
$\infty$           \\
\hline
\end{tabular}
\end{table*}

No cluster IR objects are found in clusters with either $M_{\rm i}<1.3$
M$_\odot$ or [Fe/H]$<-0.9$ (Fig.\ 13a). The L$^\prime$-band observations were
sensitive enough to have detected all AGB stars and stars on the top magnitude
of the RGB even if they had no L$^\prime$-band excess, and the old clusters
included some populous globular clusters with plenty of red giants. The
threshold for identification as an IR object is $\dot{M}_{\rm min}\sim10^{-6}$
M$_\odot$ yr$^{-1}$ (Table 8). Lower mass-loss rates, a lower dust content or
a shorter duration of the mass-loss episode (or a combination of these) might
explain the dearth of cluster IR objects with $M_{\rm i}<1.3$ M$_\odot$.

Unfortunately, there is an age-metallicity degeneracy in the sense that there
are no old metal-rich clusters or young metal-poor clusters. Galactic globular
clusters, that are all old but span a metallicity range from [Fe/H]$<-2$ to
$\sim$solar, show a clear threshold around [Fe/H]$\sim-1$ below which no
large-amplitude (Mira-type) pulsating stars are found (Frogel \& Elias 1988).
With pulsation believed to be an essential first stage in the mass-loss of red
giants, failure to pulsate strongly enough might explain the absence of
dust-enshrouded red giants at [Fe/H]$<-0.9$.

\subsection{Timescales of mass-loss}

To estimate the impact of mass loss in the superwind stage relative to the
total stellar mass, we could compare the total mass-loss rate with the total
cluster mass (Table 9). We do this for different ranges in the initial mass of
the mass-losing stars, corresponding to different ranges in cluster ages
(these are approximate as they also depend somewhat on initial metallicity).
The older clusters in our sample comprise an order of magnitude more mass than
the younger clusters, but the total mass-loss rate is much higher in the
younger clusters. This results in a very strong age dependence of the
timescale for the cluster mass to decrease, $t_{\rm mass\ loss} = \sum M_{\rm
cluster} / \sum \dot{M}$.

In principle this timescale is the time it takes for the entire cluster mass
to be dissipated if the current mass-loss rate were sustained. For stellar
systems in general (such as galaxies) it is the timescale for the feedback
mechanism to return all mass back into the ISM. It is remarkable that although
the estimated timescales span a range of more than two orders of magnitude,
they are always about 2 to 3 times the cluster age.

Another approach to quantify the timescale for mass loss is to compare the
number of cluster IR stars to the number of stars in an evolutionary stage for
which the timescale is known. For the latter we choose to count the number of
stars in our L$^\prime$-band images that are brighter than the RGB tip, which
occurs at $M_{L^\prime}=-6.4$ mag (Table 1). This has a different meaning for
clusters of different ages, for instance young clusters do not contain any RGB
stars, and will need to be calibrated against stellar evolutionary models. But
for a wide range of cluster age it does provide a rough measure for the number
of AGB stars on the thermal-pulsing AGB, whilst for the younger clusters it
provides a measure for the number of post-Main Sequence supergiants. The
latter is a consequence of our definition of the RGB tip in the
L$^\prime$-band which, coupled with large bolometric corrections for blue Main
Sequence stars discriminates between massive Main Sequence stars and redder
post-Main Sequence supergiants.

The ratio of IR objects to the total number of stars above the RGB tip is the
same within a factor two for all clusters except the oldest clusters, in which
we did not find any IR object and which anyhow contain very few stars above
the RGB tip. The ratios indicate that the superwind timescales are 10--20 per
cent of the time stars spend at $M_{L^\prime}<-6.4$ mag. There is a hint of a
trend for this timescale to be longer for older clusters (up to 20 per cent)
compared to younger clusters with AGB stars (15 per cent) and the youngest
clusters with red supergiants (10 per cent). Intermediate-mass stars spend
about $10^6$ yr above the RGB tip, which implies a timescale of
1--2$\times10^5$ yr for the superwind phase in which the stars become bright
IR objects. This corresponds to only a few thermal pulses, presumably the
final ones, and is consistent with the fact that the IR objects are found
within a magnitude from the AGB tip: this magnitude range can be fully
accounted for by luminosity variations over the thermal pulse cycle of a
factor $\sim$three (the evolutionary timescale for an AGB star to increase in
luminosity by a magnitude is about $10^6$ yr). These timescales and timing at
the end of AGB evolution confirm model predictions (e.g., Vassiliadis \& Wood
1993).

For the low- and intermediate-mass AGB stars one can estimate the importance
of the superwind phase compared to the more moderate AGB mass loss. Stars of
initial masses 1.3--1.8 M$_\odot$ will loose about 1 M$_\odot$ on the AGB. We
find 36 stars above the RGB tip in the cluster sample (Table 9), which should
therefore loose $\sim$36 M$_\odot$ in total. The ratio of this mass lost and
the integrated mass-loss rate of $1.4\times10^{-4}$ M$_\odot$ yr$^{-1}$ yields
a duration of $3\times10^5$ yr required for this mass to be lost exclusively
during the superwind phase. The same exercise for stars in the 1.8--3
M$_\odot$ range, which loose about 2 M$_\odot$ on average, yields an identical
estimate for this timescale. Given the above derived superwind timescale of
1--2$\times10^5$ yr this implies that a significant fraction of the mass loss
(30--70 per cent) indeed occurs during the superwind phase, again confirming
model predictions (e.g., Wachter et al.\ 2002).

The underlying cluster mass associated with each super-RGB-tip star increases
as the cluster ages. This can be turned around and used to estimate the
underlying cluster mass when the brightest (in L$^\prime$) cluster members can
be counted, provided that the cluster age is known. We thus obtain a first
crude estimate of such relation:
\begin{equation}
\log \frac{M[{\rm M}_\odot]}{N_{> \rm RGB}} = 3.6(\pm0.2) + 0.30(\pm0.06)\
t[{\rm Gyr}].
\end{equation}

\section{Summary of conclusions}

We present the results of a search for dusty evolved stars in star clusters in
the Magellanic Clouds. The survey was performed in first instance through
imaging in the near-IR L$^\prime$ band (at 3.8 $\mu$m) with ISAAC at the
ESO-VLT. Comparison with imaging photometry at shorter (J, H and/or K$_{\rm
s}$) and longer (MSX, IRAS, ISO and groundbased mid-IR) wavelengths allowed
the identification of objects that are surrounded by dust causing extinction
in the optical and near-IR and excess emission in the thermal IR. Targets were
selected to include populous clusters spanning a range in age and metallicity,
and clusters that were already known or suspected to be associated with
sources of bright IR emission.

Out of 9 clusters in the SMC and 29 clusters in the LMC, 19 clusters (of which
one in the SMC) were found to contain a total of 30 stars with IR excess
emission. Of these, 4 stars are probably not physically associated with the
cluster. We establish the nature of the majority of the cluster IR objects
(i.e.\ their spectral class and evolutionary state). They are all highly
evolved, and include post-superwind objects such as Planetary Nebulae. No
dusty stars are found for initial masses $M_{\rm i}<1.3$ M$_\odot$ or initial
metallicities [Fe/H]$<-0.9$. All objects in the range $1.3<M_{\rm i}<2.2$
M$_\odot$ are carbon stars, and all objects with $M_{\rm i}>6$ M$_\odot$ are
oxygen-rich giants or supergiants.

We estimate the bolometric luminosities and mass-loss rates by modelling the
spectral energy distributions with the {\sc dusty} code. The IR objects are
the most luminous cluster objects, nearly always at the maximum luminosity
expected for their initial mass and metallicity. The mass-loss rate increases
with larger progenitor mass, which may be due to a dependence on the initial
metallicity or simply the luminosity. Post-superwind objects always have the
highest mass-loss rates associated with them, where these mass-loss rates
refer to the previous superwind stage.

Stars brighter than the RGB tip in the L$^\prime$-band spend 10--20 per cent
of their time ($\sim10^5$ yr) as IR objects with mass-loss rates that exceed a
few $10^{-6}$ M$_\odot$ yr$^{-1}$. About half of the mass lost by low- and
intermediate-mass AGB stars occurs during this superwind phase.

The number of stars above the RGB tip (in L$^\prime$) can be used to estimate
the cluster mass, a relation for which we present a first crude estimate.

\begin{acknowledgements}
We thank all telescope operators who have been involved in the observations,
and the anonymous referee for her/his very positive remarks. JRM acknowledges
support through a PPARC studentship. This publication makes use of data
products from the Two Micron All Sky Survey, which is a joint project of the
University of Massachusetts and the Infrared Processing and Analysis
Center/California Institute of Technology, funded by the National Aeronautics
and Space Administration and the National Science Foundation. This research
also made use of data products from the Midcourse Space Experiment. Processing
of the data was funded by the Ballistic Missile Defense Organization with
additional support from NASA Office of Space Science. This research has also
made use of the NASA/ IPAC Infrared Science Archive, which is operated by the
Jet Propulsion Laboratory, California Institute of Technology, under contract
with the National Aeronautics and Space Administration. We also acknowledge
the use of the SIMBAD database and VizieR, operated at Centre de Donn\'{e}es
astronomiques de Strasbourg, France.
\end{acknowledgements}

\end{document}